\documentclass[10pt,journal]{IEEEtran}

\usepackage{amssymb}
\usepackage{amsmath}
\usepackage{cite}
\usepackage{url}
\usepackage{xcolor}
\usepackage{cite,graphicx,amsmath,amssymb}
\usepackage{subfigure}
\usepackage{fancyhdr}
\usepackage{mdwmath}
\usepackage{mdwtab}
\usepackage{caption}
\usepackage{amsthm}
\usepackage{setspace}
\usepackage{hyperref}
\usepackage{algorithm}
\usepackage{algorithmic}
\hypersetup{colorlinks=false}

\usepackage{array}
\usepackage{makecell}
\usepackage{threeparttable}
\usepackage{tabularx,colortbl}
\usepackage{multirow, booktabs}

\newtheorem{theorem}{Theorem}

\newtheorem{lemma}{Lemma}

\newtheorem{corollary}{Corollary}

\allowdisplaybreaks

\title{Evolution of NOMA Toward Next Generation Multiple Access (NGMA) for 6G}
\author{
        Yuanwei~Liu,~\IEEEmembership{Senior Member,~IEEE,}
        Shuowen~Zhang,~\IEEEmembership{Member,~IEEE,}
        Xidong~Mu,~\IEEEmembership{Graduate Student Member,~IEEE,}
       Zhiguo~Ding,~\IEEEmembership{Fellow,~IEEE,}
       Robert~Schober,~\IEEEmembership{Fellow,~IEEE,}
       Naofal~Al-Dhahir,~\IEEEmembership{Fellow,~IEEE,}
       Ekram~Hossain,~\IEEEmembership{Fellow,~IEEE}
       and Xuemin~Shen~\IEEEmembership{Fellow,~IEEE,}

\thanks{(Corresponding author: Yuanwei Liu)}
\thanks{Y. Liu is with the School of Electronic Engineering and Computer Science, Queen Mary University of London, London E1 4NS, U.K. (email: yuanwei.liu@qmul.ac.uk).}
\thanks{S. Zhang is with the Department of Electronic and Information Engineering, The Hong Kong Polytechnic University, Hong Kong SAR, China (e-mail: shuowen.zhang@polyu.edu.hk).}
\thanks{X. Mu is with the School of Artificial Intelligence, Beijing University of Posts and Telecommunications, Beijing 100876, China (e-mail: muxidong@bupt.edu.cn).}
\thanks{Z. Ding is with the School of Electrical and Electronic Engineering, The University of Manchester, Manchester M13 9PL, U.K. (e-mail: zhiguo.ding@manchester.ac.uk).}
\thanks{R. Schober is with the Institute for Digital Communications, Friedrich-Alexander-University Erlangen-N{\"u}rnberg (FAU), 91054 Erlangen, Germany (e-mail: robert.schober@fau.de).}
\thanks{N. Al-Dhahir is with the Department of Electrical and Computer Engineering, The University of Texas at Dallas, Richardson, TX 75080 USA (e-mail: aldhahir@utdallas.edu).}
\thanks{E. Hossain is with the Department of Electrical and Computer Engineering, University of Manitoba, Winnipeg, MB R2M 2J8, Canada (e-mail: ekram.hossain@umanitoba.ca).}
\thanks{X. Shen is with the Department of Electrical and Computer Engineering, University of Waterloo, Waterloo, ON N2L3G1, Canada (e-mail: sshen@uwaterloo.ca).}
}

\begin{document}
\maketitle
	
\begin{abstract}
Due to the explosive growth in the number of wireless devices and diverse wireless services, such as virtual/augmented reality and Internet-of-Everything, next generation wireless networks face unprecedented challenges caused by heterogeneous data traffic, massive connectivity, and ultra-high bandwidth efficiency and ultra-low latency requirements. To address these challenges, advanced multiple access schemes are expected to be developed, namely next generation multiple access (NGMA), which are capable of supporting massive numbers of users in a more resource- and complexity-efficient manner than existing multiple access schemes. As the research on NGMA is in a very early stage, in this paper, we explore the evolution of NGMA with a particular focus on  non-orthogonal multiple access (NOMA), i.e., the transition from NOMA to NGMA. In particular, we first review the fundamental capacity limits of NOMA, elaborate on the new requirements for NGMA, and discuss  several possible candidate techniques. Moreover, given the high compatibility and flexibility of NOMA, we provide an overview of current research efforts on multi-antenna techniques for NOMA, promising future application scenarios of NOMA, and the interplay between NOMA and other emerging physical layer techniques. Furthermore, we discuss advanced mathematical tools for facilitating the design of NOMA communication systems, including conventional optimization approaches and new machine learning techniques. Next, we propose a unified framework for NGMA based on multiple antennas and NOMA, where both downlink and uplink transmissions are considered, thus setting the foundation for this emerging research area. Finally, several practical implementation challenges for NGMA are highlighted as motivation for future work.
\end{abstract}

\begin{IEEEkeywords}
6G, next generation multiple access (NGMA), non-orthogonal multiple access (NOMA), multi-antenna techniques, mathematical optimization, machine learning.
\end{IEEEkeywords}	

\section{Introduction}
By the end of 2023, Cisco predicts that the number of mobile users will reach 13.1 billion and the number of Internet-enabled devices will have increased from 18.4 billion in 2018 to 29.3 billion~\cite{cisco2020cisco}. This explosive demand will gradually overwhelm the connectivity capabilities of the fourth generation (4G) and fifth generation (5G) wireless networks. Additionally, this increasing trend shows no sign of slowing down in the next decade since this demand is caused only by around 70\% of the global population. Therefore, providing massive access is one of the most important targets for next generation wireless networks, namely the sixth-generation (6G). Currently, the research on 6G is being carried out on the global stage, e.g., the ``6G Flagship'' in Finland~\cite{Finland}, ``6G Hubs'' in Germany, ``Broadband Communications and New Networks'' in China, Terahertz (THz) communication studies in the USA, to name a few. To support future Internet of Everything (IoE) and Mobile Internet, 6G has several critical performance targets~\cite{8766143}: a) The peak data rate needs to be at least one terabit per second; b) the air interface latency shall range from 0.01 to 0.1 millisecond for users with high mobility; c) the connectivity density shall be 10 times larger than in current 5G systems; d) the spectral efficiency (SE) and energy efficiency (EE) have to be 5-10 and 10-100 times higher than for 5G, respectively; and e) the reliability has to be higher than 99.99999\%~\cite{han2019terahertz}.\\
\indent In order to achieve the aforementioned stringent targets, one of the most fundamental issues is the design of sophisticated multiple access techniques for the next generation wireless networks, namely next generation multiple access (NGMA). The goal of NGMA is to enable a tremendous number of users/devices to be efficiently and flexibly connected with the network over the given wireless radio resources. Looking back at the development history of wireless communications, multiple access techniques have been key enablers. Depending on whether each user is allocated orthogonal resources or not, the current multiple access techniques can be loosely classified into two categories, namely orthogonal transmission strategies and non-orthogonal transmission strategies~\cite{Yuanwei2017pieee,Ding_survey}. In particular, given their advantages of low complexity and interference-avoidance, orthogonal transmission strategies have been extensively employed in practical wireless communication systems, such as frequency division multiple access (FDMA) in the first generation (1G), time division multiple access (TDMA) in the second generation (2G)~\cite{TDMA}, code division multiple access (CDMA) with orthogonal codes in the third generation (3G)~\cite{CDMA}, and orthogonal frequency division multiple access (OFDMA) in the 4G~\cite{OFDMA}, where users are allocated with orthogonal frequency/time/code resource blocks. However, given the ever increasing number of wireless devices and the limited amount of available spectrum, conventional orthogonal transmission strategies have become inefficient due to their low SE and the limited number of users/devices that can be supported due to the inflexible resource allocation. As a result, to handle the challenging emerging heterogeneous services and applications (e.g., virtual reality (VR), augmented reality (AR), and industry 4.0), non-orthogonal transmission strategies are widely considered to be a promising solution and have drawn significant research interests in the past few years~\cite{Yuanwei2017pieee,Ding_survey,Zhiguo_mag,Dai_survey,Open_survey}. Different from orthogonal transmission strategies, the key idea of non-orthogonal transmission strategies is to allow different users to share the same resource blocks. To deal with the resulting interference caused by the non-orthogonal resource allocation, additional techniques have to be employed at the transmitters and receivers~\cite{Fundamentals}, including but not limited to superposition coding (SC), rate splitting (RS), successive interference cancellation (SIC), and message passing (MP). Although the non-orthogonal transmission strategies increase the transmitter and receiver complexity, significant benefits can be achieved, such as supporting massive connectivity~\cite{Yuanwei_Large-Scale,Liu_Large}, achieving high SE and EE~\cite{Zhiguo_random,yuanwei_SWIPT}, and guaranteeing user fairness~\cite{16_CL_Liu}. As a result, compared with orthogonal transmission strategies, the aforementioned benefits make non-orthogonal transmission strategies promising candidates for NGMA. Nevertheless, the investigation of NGMA is still in an infancy stage. Besides achieving high bandwidth efficiency and supporting high connectivity, NGMA has to facilitate diverse application scenarios and be compatible with other promising physical layer techniques.\\
\begin{table}[htb]\footnotesize
\caption{List of Acronyms}
\centering
\begin{tabular}{|l||l|}
\hline
A2G& Air-to-Ground \\
AP&Access Point\\
AWGN & Additive White Gaussian Noise \\
AR & Augmented Reality\\
BC & Broadcast Channel\\
CDMA & Code Division Multiple Access \\
CSI & Channel State Information\\
BS & Base Station\\
DL& Deep Learning\\
DoF& Degree-of-Freedom\\
DPC&Dirty-Paper Coding\\
EE & Energy Efficiency\\
FD& Full Duplex\\
FDMA & Frequency Division Multiple Access\\
FL&Federated Learning\\
GB& Grant-Based\\
GF & Grant-Free\\
IRS & Intelligent Reflecting Surface\\
ISaC & Integrated Sensing and Communication\\
IoE & Internet of Everything \\
IoT & Internet of Things\\
KPI &  Key Performance Indicator\\
LoS & Line-of-Sight\\
MAC &  Multiple-Access Channel\\
MC-MTC&Massive and Critical Machine-Type Communication \\
mmWave & Millimeter-wave\\
MnAC & Many-Access Channel \\
MEC & Mobile Edge Computing\\
MIMO & Multiple-Input Multiple-Output\\
MISO & Multiple-Input Single-Output\\
ML & Machine Learning\\
MP & Message Passing\\
MUD & Multiuser Detection\\
NGMA & Next Generation Multiple Access\\
NOMA & Non-Orthogonal Multiple Access\\
OFDMA & Orthogonal Frequency Division Multiple Access \\
OMA & Orthogonal Multiple Access\\
OTFS&Orthogonal Time Frequency Space\\
PDMA &  Pattern Division Multiple Access\\
QoS& Quality of Service\\
RF& Radio Frequency \\
RIS& Reconfigurable Intelligent Surface\\
RL & Reinforcement Learning\\
RS & Rate Splitting\\
RSMA & Rate Splitting Multiple Access\\
SE & Spectral Efficiency\\
SC & Superposition Coding \\
SCA & Successive Convex Approximation\\
SCMA& Sparse Code Multiple Access\\
SDMA& Space Division Multiple Access\\
SIC& Successive Interference Cancelation\\
SIMO& Single-Input Multiple-Output\\
SINR& Signal-to-Interference-plus-Noise Ratio\\
SISO& Single-Input Single-Output\\
TDMA & Time Division Multiple Access\\
THz & Terahertz\\
UAV & Unmanned Aerial Vehicle\\
VLC & Visible Light Communication\\
VR& Virtual Reality\\
ZF&Zero-Forcing\\
\hline
\end{tabular}
\label{table:abbre}
\end{table}



\indent In the above context, the goal of this paper is to set the foundation for this new NGMA research for academic and industrial researchers. In particular, this paper investigates the development of NGMA with a particular focus on exploiting power-domain NOMA\footnote{Unless stated otherwise, NOMA refers to ``power-domain NOMA'' in this paper.}, i.e., the transition from NOMA to NGMA. In Section II, we first review the information-theoretic capacity limits of NOMA, and discuss the requirements that NGMA schemes have to fulfill and some possible candidates based on different non-orthogonal transmission strategies. Focusing on NOMA, we provide an overview of the current research contributions on multi-antenna NOMA, promising application scenarios for NOMA in 6G, and the interplay between NOMA and other emerging physical layer techniques in Sections III, IV, and V, respectively. In Section VI, advanced mathematical tools, including both conventional optimization and new machine learning (ML) methods, that support the design of NOMA communication systems and help uncover their full benefits are discussed. In Section VII, we propose a unified framework for NGMA based on multi-antenna techniques and NOMA for both downlink and uplink transmission. In Section VIII, we highlight several implementation challenges and future work for NGMA. Finally, Section IX concludes the paper. Table \ref{table:abbre} provides a list of acronyms used in this paper.
\section{Requirements for NGMA Design and Possible Candidates}
To start with, we first review the fundamental information-theoretic capacity limits of NOMA to highlight the basic benefits of NOMA. Then, to cater to the new requirements in next generation cellular networks, we identify several new challenging issues that need to be considered, and which motivate new lines of research to unveil the fundamental limits of NGMA and the design of practical schemes for approaching them. Moreover, some possible candidates for NGMA exploiting different non-orthogonal transmission strategies are introduced.

\subsection{Information-Theoretic Capacity Limits of NOMA}
In this subsection, we discuss the information-theoretic capacity limits of NOMA in both single-antenna and multi-antenna communication systems.
\begin{figure*}[t]
\begin{center}
    \includegraphics[width=0.8\linewidth]{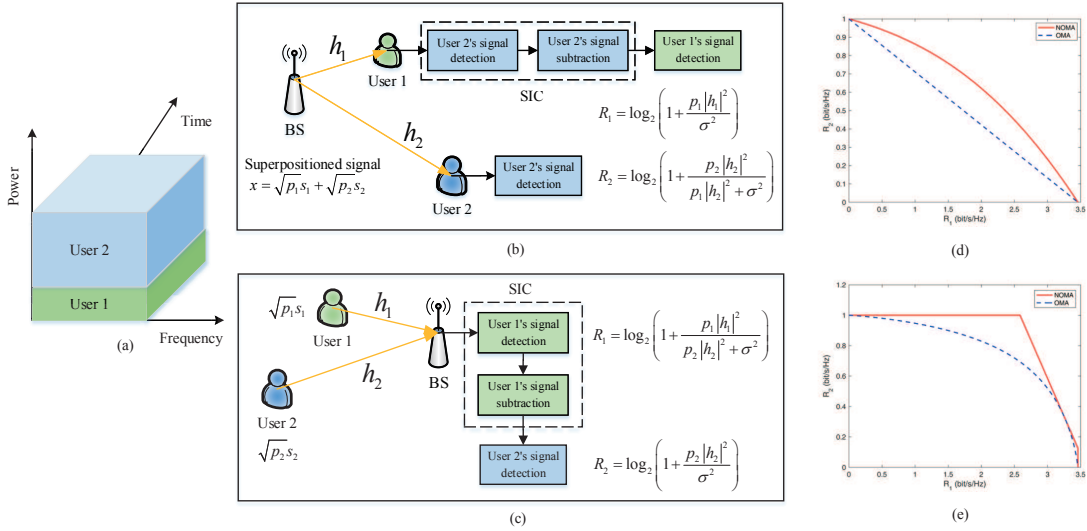}
    \caption{Illustration of two-user NOMA. (a) The signals of two users are multiplexed in the power domain using the same time/frequency resource. (b) Downlink NOMA transmission. (c) Uplink NOMA transmission. (d) BC capacity/rate region comparison, $0\le p_1+p_2\le 1$W. (e) MAC capacity/rate region comparison, $0\le p_1\le 1$W, $0\le p_2\le 1$W. We set $\frac{{{{\left| {{h_1}} \right|}^2}}}{{{\sigma ^2}}} = 10$ dB and $\frac{{{{\left| {{h_2}} \right|}^2}}}{{{\sigma ^2}}} = 0$ dB, where $h_k$ and $p_k$ denote the channel coefficient and transmit power of user $k \in \left\{ {1,2} \right\}$, respectively, ${\sigma ^2}$ denotes the noise power at both users, and $R_k$ denotes the communication rate achieved by user $k \in \left\{ {1,2} \right\}$.}
    \label{capacity}
\end{center}
\end{figure*}

\subsubsection{Single-Input Single-Output (SISO) case} For single-antenna transmitters and receivers, orthogonal multiple access (OMA) schemes have been widely employed in pervious generations of cellular networks, including FDMA in 1G, TDMA in 2G, CDMA in 3G, and OFDMA in 4G long-term evolution (LTE), where users are served through orthogonal time/frequency/code resources. Despite the advantage of completely avoiding inter-user interference, the achievable rate regions of OMA for both the Gaussian SISO multiple-access channel (MAC) and broadcast channel (BC) are known to be a subset of the corresponding capacity region in general \cite{Elements}, since each user can use only a portion of the available resource blocks. As a result, from an information-theoretic perspective, OMA schemes are strictly suboptimal transmission strategies. In contrast, it has been widely recognized that NOMA is capable of achieving the capacity region in the SISO case. In NOMA, the users share the same resources\footnote{Here, the term ``resources'' includes the time, frequency, and code resources as well as combinations thereof.}, and are separated in the \emph{power} domain with the aid of SC and SIC. The two-user power multiplexing of NOMA is illustrated in Fig. \ref{capacity}(a). For the two-user downlink NOMA transmission in Fig. \ref{capacity}(b), the two users' signals are superimposed via SC and transmitted by the BS. Assume that user 1 has better channel conditions than user 2. User 1 first decodes user 2's signal and subtracts it from the received signal via SIC. Then, it decodes its intended signal, while user 2 directly decodes its intended signal by treating user 1's signal as noise. For the two-user uplink NOMA transmission in Fig. \ref{capacity}(c), the two users send their signals to the BS using the same resources. At the BS, the two users' signals are successively decoded for a given decoding order via SIC. Examples of comparisons between the NOMA capacity region and the OMA rate region for the single-antenna BC and MAC are depicted in Fig. \ref{capacity}(d) and Fig. \ref{capacity}(e), respectively. In both instances, NOMA is able to achieve the capacity region~\cite{72_TIT_Cover}, while OMA is only able to achieve specific points on the capacity region.

\subsubsection{Multiple-Input Single-Output (MISO)/Multiple-Input Multiple-Output (MIMO) case} One key feature of next generation wireless networks is that multiple antennas will be installed at the transmitters and/or the receivers, which requires research efforts to be devoted to investigating the capacity region of multi-user MISO/MIMO systems. For the general Gaussian MISO/MIMO MAC, it was shown that the capacity region can also be achieved employing SIC~\cite{1262622,1237143}, i.e., NOMA is still capacity-achieving.\\
\indent However, due to the non-degraded nature of the general Gaussian MISO/MIMO BC, the capacity region is not known in general whereas the sum-rate capacity was shown to be achievable with the dirty paper coding (DPC)~\cite{1237143,1207369}. Moreover, it was shown that the DPC rate region coincides with the capacity region of the Gaussian MIMO BC in some special cases, i.e., the aligned and degraded MIMO BC (ADBC) and the aligned MIMO BC (AMBC)~\cite{1683918}. Therefore, although the high complexity of DPC makes a practical implementation challenging, the DPC rate region provides a useful performance benchmark for other multi-user MISO/MIMO transmission schemes. In general, the DPC rate region is larger than the NOMA rate region for MISO/MIMO systems. Take the two-user MISO BC as an example, where the two users are assumed to be indexed by user 1 and user 2. The main reason for the rate gain of DPC over NOMA is that for a given user ordering, e.g., from user 2 to user 1, for DPC, the BS first encodes the information intended for user 2, and then encodes the information intended for user 1 by pre-subtracting user 2's message, i.e., the inter-user interference is pre-cancelled by the BS. By contrast, for NOMA, user 1 employs SIC to cancel the interference caused by user 2's signal, which imposes additional constraints to ensure user 2's signal received at user 1 has a sufficient power such that SIC can be successfully carried out. As a result, DPC generally outperforms NOMA in terms of rate. Nevertheless, in some special cases, NOMA can achieve the same performance as DPC. For instance, the authors of \cite{quasi-degradation} derived a sufficient and necessary condition for the MISO BC, namely the quasi-degradation condition for NOMA to approach the DPC rate region. For the MIMO BC, the authors of \cite{1683918} revealed that the capacity region of the ADBC, the DPC rate region, and the NOMA rate region are identical. It is worth noting that although NOMA is not always capacity-achieving for the general Gaussian MISO/MIMO BC, it is still an effective transmission strategy, especially for the overloaded regime\footnote{Given the rapidly growing number of users/devices, next generation wireless networks are expected to encounter more overloaded use cases (e.g., IoE and massive machine type communication (mMTC)) than previous and current wireless networks, where underloaded use cases are more common.}~\cite{16_TWC_Ding2,16_TWC_Ding}, where the number of available antennas at the BS is smaller than the total number of user antennas, as will be discussed in Section VII.A.

\subsection{New Considerations for NGMA Design}
In this subsection, we identify several new considerations that need to be taken into account for NGMA design.
\subsubsection{Massive access}
With the increasing demand for expanding cellular connectivity as well as the prosperous Internet of Things (IoT) market, the next generation cellular networks will be characterized by an excessively large number of users, a scenario also referred to as \emph{massive connectivity} or \emph{massive access} \cite{18_TSP_Liu}. The existing information-theoretic results on multiple access schemes mostly focus on finding the capacity limits for an infinite coding blocklength and a fixed number of users. However, for massive access channels where the number of users grows with the blocklength, the fundamental limits need to be revisited. To this end, the authors of \cite{17_TIT_Chen} proposed the notion of \emph{many-access channel (MnAC)} for modeling a channel with a single receiver and multiple transmitters, the number of which grows unboundedly with the blocklength. The new notions of achievable message length and symmetric capacity were proposed in \cite{17_TIT_Chen}, where separate identification and decoding were shown to be capacity achieving. These new results will serve as valuable performance upper bounds for NGMA in the many-user asymptotic regime.

\subsubsection{Short-packet transmission}
Besides massive access, another new feature that needs to be considered for NGMA is \emph{short-packet transmission} in IoT applications, where the coding blocklength is finite and short due to the low latency requirement \cite{16_TCOM_Durisi}. For this regime, the authors of \cite{10_TIT_Polyanskiy} derived the maximum channel coding rate for a given blocklength and error probability. Particularly, the rate gap compared to the case with infinite coding blocklength was quantified. How to extend this result to the regime with a large yet finite number of users and a finite coding blocklength is still an open problem \cite{17_TIT_Chen}. Moreover, approaching the fundamental limits with practical transmission and multiple access schemes requires future research efforts.

\subsection{Possible Candidates for NGMA}

As discussed in the previous subsection, the trend of NGMA is to transition from orthogonality to non-orthogonality, i.e., serving multiple users/devices by allowing them to share the same resources instead of allocating them to dedicated orthogonal resources. The reason behind this trend can be explained as follows. On the one hand, orthogonal transmission schemes are strictly suboptimal compared to non-orthogonal transmission schemes. On the other hand, given the explosively growing number of users/devices in next generation wireless networks, orthogonal transmission schemes can only support a limited number of users/devices for given available orthogonal resources. This calls for the development of non-orthogonal transmission schemes for NGMA. In the following, based on the concept of non-orthogonality, some possible candidates for NGMA are introduced along with their technical foundations.

\subsubsection{Power-Domain NOMA (PD-NOMA)} The key idea of PD-NOMA is to serve multiple users in the same time/frequency/code resources and distinguish them in the power domain~\cite{Yuanwei2017pieee,Ding_survey}. SC and SIC are the two key technologies in PD-NOMA, which have been proven to be capacity-achieving in the single-antenna BC and MAC. The basic principle has been introduced in Section II.A.1 for the two-user case. For broadband communications over frequency-selective fading channels, where the channel coherence bandwidth is smaller than the system bandwidth, PD-NOMA can be straightforwardly integrated with OFDMA by assigning multiple users to each OFDMA subcarrier and serving them with PD-NOMA. This approach was adopted in multiuser superposition transmission (MUST)~\cite{MUST}, which was incorporated into LTE-A for simultaneously supporting two users on the same OFDMA subcarrier. Another application of PD-NOMA is layered division multiplexing (LDM)~\cite{LDM}, which was included in the digital TV standard (ATSC 3.0) to delivery multiple superpositioned data streams for TV broadcasting.

\subsubsection{Code-Domain NOMA (CD-NOMA)} Inspired by CDMA, where multiple users are served via the same time/freqency resources and distinguished by the allocated dedicated user-specific spreading sequences, CD-NOMA was proposed, whose key idea is still to serve multiple users in the same time/freqency resources but employing user-specific spreading sequences which are either sparse sequences or non-orthogonal cross-correlation sequences having low cross-correlation~\cite{Dai_survey}. At the receiver, multiuser detection (MUD) is usually carried out in an iterative manner using MP based algorithms. The family of CD-NOMA schemes has many members, such as Low-Density Signature (LDS)-CDMA~\cite{LDS-CDMA}, LDS-OFDM~\cite{LDS-OFDM}, Sparse Code Multiple Access (SCMA)~\cite{SCMA}, and Pattern Division Multiple Access (PDMA)~\cite{PDMA}.

\subsubsection{Space Division Multiple Access (SDMA)} With the rapid development of multi-antenna techniques, MIMO communication plays an important role in the current 5G standard and the upcoming future wireless networks. By installing multiple antennas at transmitters and/or receivers, additional spatial DoFs can be exploited as compared to single-antenna communication systems~\cite{SDMA0}. As a result, multiple users/devices in SDMA can be served in the same time/frequency/code domain and distinguished in the spatial domain. The most commonly employed method for SDMA is linear precoding (LP) due to its low complexity. In particular, the inter-user interference can be effectively mitigated by exploiting the spatial DoFs to design suitable transmit and/or receive beamformers. However, SDMA is only applicable in the underloaded and critically loaded regimes as the available spatial DoFs are exploited for mitigating inter-user interference.

\subsubsection{Rate Splitting Multiple Access (RSMA)} RSMA is a new non-orthogonal multi-antenna transmission scheme, which was proposed in recent years for multi-user multi-antenna communications. Relying on the rate splitting technique~\cite{RSMA}, the transmitter splits part of each user's message (i.e., the private message) into a common message, which is intended for all served users. The remaining private messages and the constructed common message are transmitted via beamformers, as in SDMA. The receivers treat the private messages of all users as interference when decoding the common message and subtract the common message from the received signal (i.e., SIC). Subsequently, the intended private message is decoded and combined with part of the decoded common message. Therefore, compared to SDMA, RSMA provides more flexibility in mitigating interference~\cite{RSMA2}.\\
\indent Given the aforementioned possible candidates, in this paper, we focus our attention on exploiting NOMA to develop NGMA~\cite{liu2021application}. The main reasons can be summarized as follows. On the one hand, we expect the overloaded regime to be an important use case for next generation wireless networks, for which NOMA is a promising candidate. On the other hand, the existing research contributions have shown that NOMA provides a higher degree of compatibility and flexibility. This enables the synergistic integration of NOMA with other components of next generation networks, such as multi-antenna techniques, challenging application scenarios, and other physical layer techniques, as will be explained in detail in the following.
\section{Multi-Antenna Techniques for NOMA}\label{sec_multi_antenna}
Multi-antenna techniques are envisioned to be an indispensable component of NGMA. By equipping the BS and/or users with multiple antennas, additional DoFs in the \emph{spatial} domain can be exploited for enhancing NGMA performance compared to the case with single-antenna BSs and users. In this section, we first review the existing multi-antenna techniques developed for NOMA (also called ``MIMO-NOMA''), and then discuss several promising directions of future NOMA multi-antenna techniques towards NGMA.
\begin{figure*}[t!]
\begin{center}
    \includegraphics[width=6in]{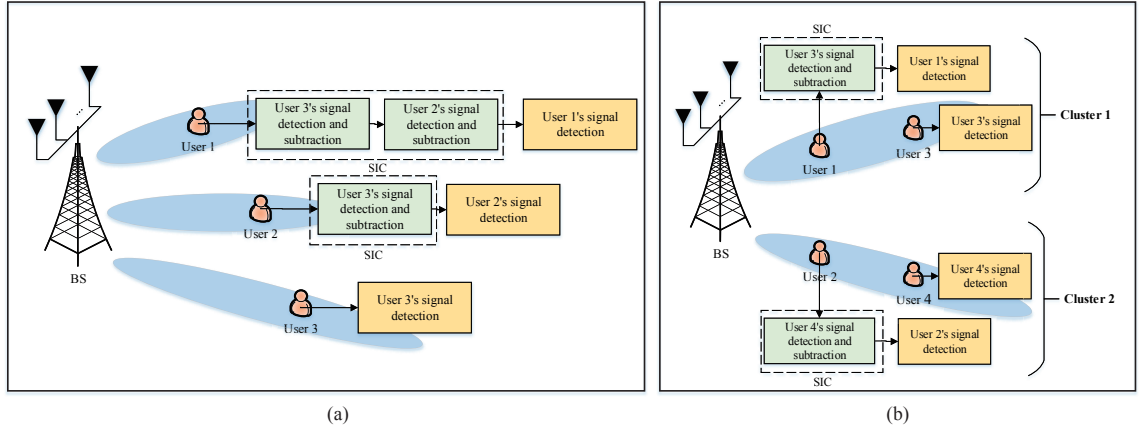}
    \caption{Illustration of existing multi-antenna techniques for NOMA. (a) Beamformer-based MIMO-NOMA with 3 users using 3 beamformers. (b) Cluster-based MIMO-NOMA with 4 users in 2 clusters using 2 beamformers~\cite{18_WC_Liu}.}
    \label{BBCB}
\end{center}
\end{figure*}
\subsection{Review of Existing NOMA Multi-Antenna Techniques}
Compared to single-antenna based NOMA (i.e., SISO-NOMA) where both the BS and users are equipped with a single antenna, an important new design aspect in MIMO-NOMA is the transmit/receive \emph{beamformer} at the BS and/or users \cite{18_WC_Liu,18_WC_Huang}. On the one hand, similar to the case of MIMO-OMA, the beamformer design critically determines the signal power as well as interference power at the different users, which thus plays an important role for the user's signal-to-interference-plus-noise ratio (SINR). On the other hand, the SIC performance in MIMO-NOMA critically depends on the decoding order of the users' messages, which needs to be judiciously designed jointly with the beamformer. In SISO-NOMA, the decoding order can be easily optimized by ordering the user channel gains, since the signal and interference powers at each user are only determined by the scalar BS-user channels. In contrast, the signal and interference powers at MIMO-NOMA users are jointly determined by the BS-user channel matrices (or vectors) and the transmit/receive beamformers at the BS/users. This leads to a new class of joint optimization problems for the beamformers and resource allocation (e.g., power/sub-channel allocation) that are more challenging than the corresponding design problems for MIMO-OMA and SISO-NOMA. Moreover, the impact of the beamformer design on the performance of MIMO-NOMA also requires new analysis techniques. In the literature, tremendous research efforts have been devoted to studying the optimization and analysis of MIMO-NOMA systems. Specifically, two types of beamforming strategies have been proposed for MIMO-NOMA, namely, \emph{beamformer-based MIMO-NOMA} and \emph{cluster-based MIMO-NOMA} \cite{18_WC_Liu}, as detailed below.	

\subsubsection{Beamformer-based MIMO-NOMA}
Similar to SDMA, a straightforward transmit/receive beamforming strategy in MIMO-NOMA is to design a linear beamformer for each user,\footnote{For the purpose of exposition, we focus on the case where each user only has one data stream.} where new constraints need to be added for ensuring the efficacy of SIC\footnote{This issue will be discussed in detail in Section VII.A.}. An example of 3-user downlink beamformer-based MIMO-NOMA communication is illustrated in Fig. \ref{BBCB}(a). Different from SDMA, where the BS beamformers need to maximize the desired message power at each user while suppressing the inter-user interference, the BS beamformers in MIMO-NOMA should ensure that the received power of a message is strong for both the intended user and all the other ``stronger'' users that need to decode this message before their own messages in SIC. Therefore, new methodologies need to be developed for meeting the drastically different requirements of MIMO-NOMA compared to SDMA. It is worth noting that although beamformer-based MIMO is inspired by SDMA, it is applicable to both underloaded/critically loaded and overloaded regimes~\cite{16_TSP_Hanif,quasi-degradation}.\\
\indent Along this line, considering a two-user downlink MIMO-NOMA system with a given decoding order, the authors of \cite{15_WCL_Sun} investigated the transmit covariance matrix optimization problem to maximize the ergodic capacity based on statistical channel state information (CSI). Moreover, motivated by the low complexity of the horizontal Bell Labs layered space-time (H-BLAST) scheme in conventional MIMO, the authors of \cite{16_TWC_Choi} proposed a MIMO-NOMA scheme with layered transmission, and studied the power allocation optimization for sum-rate maximization. The authors of \cite{16_TSP_Hanif} considered a multi-user downlink MISO-NOMA system with given decoding order, and developed a minorization-maximization algorithm for the beamformer design to maximize the sum rate. The numerical results in \cite{16_TSP_Hanif} suggest that MIMO-NOMA may outperform the conventional Zero-Forcing (ZF) based beamforming when the number of users is much larger than the number of transmit antennas at the BS. The authors of \cite{quasi-degradation} identified a so-called ``quasi-degraded'' channel condition for the two-user MISO broadcast channel. Under this condition, the optimal MISO-NOMA beamforming strategy for minimizing the transmit power under user rate constraints was derived, and was shown to achieve the same performance as the optimal DPC based scheme.		
\subsubsection{Cluster-based MIMO-NOMA}
In beamformer-based MIMO-NOMA, each user's message is interfered by all the other users' messages in the network. Therefore, jointly optimizing the decoding order and the beamformer requires exponential complexity with respect to the number of users, which is especially prohibitive when the number of users is very large, i.e., the system is extremely overloaded. To circumvent this issue, a cluster-based MIMO-NOMA scheme was proposed~\cite{16_CL_Liu}, where the users are grouped into multiple clusters, with all users in a cluster sharing the same beamformer. An example of 4 users grouped into 2 clusters in downlink cluster-based MIMO-NOMA communication is illustrated in Fig. \ref{BBCB}(b). The advantages of cluster-based MIMO-NOMA can be summarized as follows. On the one hand, users with similar spatial features can be grouped into the same cluster, which enables the inter-cluster interference to be significantly suppressed or even eliminated by carefully designing the beamformers for the different clusters. As a result, the limited spatial DoFs can be exploited in a more efficient manner, which is of vital importance for the overloaded regime. On the other hand, as SIC only needs to be carried out for users within the same cluster, the resulting decoding order among intra-cluster users can be more easily determined due to the reduced number of users that need to be considered.\\
\indent Due to the appealing features outlined above, cluster-based MIMO-NOMA has attracted significant research interests. Specifically, a straightforward approach to design the beamformers for each cluster is to apply the ZF criterion, so as to completely remove the inter-cluster interference (see, e.g., \cite{16_CL_Liu,16_TWC_Ding2,16_TWC_Ding,15_TCOM_Choi,17_WCL_Zeng,18_Access_Zeng}). Based on the ZF beamforming design, the authors of \cite{16_CL_Liu} proposed efficient user clustering algorithms for downlink MIMO-NOMA for maximization of the fairness among the users, and further designed the power allocation for the users in each cluster. The authors of \cite{16_TWC_Ding2} proposed a BS precoding and user detection design for a MIMO-NOMA downlink system, and analyzed its performance in terms of the outage probability and diversity order. The effect of user clustering on the performance was also investigated. Moreover, by leveraging a signal alignment technique, a general MIMO-NOMA framework for both downlink and uplink transmission was proposed in \cite{16_TWC_Ding}. This scheme is able to transform the MIMO-NOMA system into several parallel SISO-NOMA systems with lower design complexity. Its performance was analyzed by leveraging stochastic geometry, where the effect of various power allocation policies on the performance was also revealed. Furthermore, the authors of \cite{15_TCOM_Choi} considered a two-user downlink MISO-NOMA system and proposed a two-stage beamforming scheme, where the first stage performs ZF beamforming for nulling the inter-cluster interference, and the second stage employs intra-cluster beamformers for the users in the same cluster in order to minimize the transmit power. The authors of \cite{17_WCL_Zeng} compared the sum achievable rates of MIMO-NOMA with user clustering and MIMO-OMA, where the superior performance of the optimized MIMO-NOMA scheme was revealed. Under user quality-of-service (QoS) constraints, the authors of \cite{18_Access_Zeng} investigated the power allocation and user admission optimization problem to maximize the EE in a multi-cluster MIMO-NOMA downlink system.\\
\indent On the other hand, several works have considered different beamformer designs that tolerate a certain level of inter-cluster interference. Along this line, the authors of \cite{17_Access_Ali} proposed a beamforming technique for the MIMO-NOMA downlink that is able to cancel a significant portion of the inter-cluster interference when the number of BS transmit antennas is smaller than the total number of antennas at the users. Moreover, an efficient user clustering method was proposed in \cite{17_Access_Ali} that aims to allocate users with maximally distinct channel gains to each cluster in order to maximize the SIC performance. Considering two different types of imperfect CSI knowledge, the authors of \cite{18_TWC_Cui2} studied the beamformer optimization problem under outage probability constraints, and proposed efficient algorithms based on  successive convex approximation (SCA) and semi-definite programming (SDP) techniques.
\subsection{Future Multi-Antenna Techniques for NOMA}
Besides the above works that investigated MIMO-NOMA for general communication systems, next generation cellular networks require the synergistic interplay of MIMO-NOMA and various other new technologies such as massive MIMO and millimeter-wave (mmWave)/THz communications. In the following, we discuss several promising directions for the joint design of multi-antenna techniques for MIMO-NOMA and these technologies.
\subsubsection{Massive MIMO-NOMA}
Massive MIMO has been recognized as a key technology to boost the SE in next generation cellular networks due to the large number of spatial DoFs introduced by the massive antennas equipped at the BS \cite{14_CM_Larsson}. Traditional massive MIMO literature has mostly focused on the underloaded regime, where the number of users is smaller than the number of antennas at the BS. However, considering the future need of connecting an excessive number of users in cellular networks (i.e., the overloaded regime), the spatial DoFs provided by massive MIMO may not be sufficient. In this case, NOMA can help to serve more users via power-domain multiplexing. Along this line, the authors of \cite{19_TWC_Liu} devised a Gaussian message passing (GMP) MUD algorithm for overloaded massive MIMO-NOMA systems, whose convergence was carefully analyzed. Motivated by the new cell-free massive MIMO architecture featuring a large number of spatially distributed BSs without cell boundaries, the authors of \cite{20_CL_Rezaei} developed a low-complexity user clustering method for cell-free massive MIMO-NOMA underlaid below a primary massive MIMO system, and characterized its achievable sum rate. The authors of \cite{19_JSTSP_Zeng} analyzed the secrecy performance of an artificial noise (AN) aided massive MIMO-NOMA network, where the results showed that the proposed system outperforms conventional massive MIMO-OMA systems in terms of secrecy rate and EE.\\
\indent Despite the promises it holds, massive MIMO-NOMA also faces new design challenges. Specifically, as discussed above, beamformer design is critical for the performance of MIMO-NOMA, which requires accurate CSI knowledge. However, in massive MIMO, acquisition of CSI requires substantial channel training and feedback overhead, especially for frequency-division duplex (FDD) systems. To resolve this issue, the authors of \cite{16_SPL_Ding} proposed and analyzed a massive MIMO-NOMA scheme for the case of limited feedback bits, by decomposing the system into multiple SISO-NOMA systems. The authors of \cite{17_JSAC_Ma} proposed an iterative data-aided channel estimation scheme based on an orthogonal pilot structure and a superimposed pilot structure, where the NOMA concept was shown to improve the channel estimation performance and alleviate the pilot contamination problem. Moreover, to enjoy favorable propagation conditions, the antennas of massive MIMO systems need to be spaced by at least half a wavelength so that their channels are independent of each other. To reduce the size of the antenna array, the authors of \cite{19_TWC_Sena} considered a massive MIMO-NOMA design with dual-polarized antennas, where the antennas with orthogonal polarity can be co-located without causing channel correlation. Two beamforming schemes were proposed for the considered system, and the corresponding outage probabilities were derived in closed-form. The effect of imperfect SIC on the outage probability and ergodic rate of massive MIMO-NOMA was analyzed in \cite{20_TWC_Sena2}. To combat the aforementioned channel estimation/SIC imperfections and to improve the reliability of massive MIMO-NOMA, the authors of \cite{20_TWC_Sena} devised a successive sub-array activation (SSAA) diversity scheme, whose outage probability was characterized in closed form. The proposed scheme was shown to outperform the conventional massive MIMO-NOMA.

\subsubsection{MIMO-NOMA at mmWave/THz bands}
Besides massive MIMO, another promising approach for boosting the data rate in next generation cellular networks is by utilizing the huge bandwidths offered in the mmWave or even THz frequency bands. However, a key design challenge for mmWave/THz communications lies in the significant hardware complexity and cost at high frequencies. To overcome this issue, a \emph{beamspace MIMO} technique was proposed, where the MIMO channel is transformed into sparse beamspace channels via lens antenna arrays to reduce the required number of radio frequency (RF) chains~\cite{13_TAP_Brady}. Nevertheless, the number of users served by beamspace MIMO cannot exceed the number of RF chains, which thus motivates the adoption of NOMA to increase the number of supportable users by exploiting the power-domain multiplexing gain~\cite{8635489}. Along this line, the authors of \cite{17_JSAC_Wang} proposed a novel beamspace MIMO-NOMA scheme with efficient ZF beamforming design and dynamic power allocation, which was shown to significantly outperform existing beamspace MIMO systems that do not leverage NOMA. The authors of~\cite{20_JSAC_Tang} developed a beam selection algorithm for maximization of the sum-rate of beamspace MIMO-NOMA, and further proposed a detection algorithm with low complexity. The authors of \cite{20_TSP_Jiao} proposed an alternating optimization algorithm for optimizing the power allocation and beamformers to maximize the minimum rate of a beamspace MIMO-NOMA system.\\
\indent Aside from beamspace MIMO, another technique for reducing the number of RF chains needed is \emph{hybrid precoding}~\cite{14_JSTSP_Heath}, which relies on a small-size digital precoder that needs a small number of RF chains and is connected to a larger-size analog precoder. Along this direction, the authors of \cite{19_JSAC_Dai} considered hybrid precoding for mmWave MIMO-NOMA system with simultaneous wireless information and power transfer (SWIPT), and proposed the joint design of user grouping, hybrid precoding, power allocation, and power splitting for SWIPT. Taking into account angular estimation errors, the authors of \cite{19_JSTSP_Hu} analyzed the achievable rate of a mmWave MIMO-NOMA system employing hybrid precoding, and proposed a novel cluster grouping algorithm for suppressing the inter-cluster interference. The authors of \cite{20_JSAC_Zhang} studied a THz MIMO-NOMA system and optimized the user clustering, hybrid precoding, and power allocation to maximize the EE.
\section{Promising Application Scenarios of NOMA towards 6G}
Having reviewed the current research progress regarding multi-antenna techniques in NOMA, in this section, we discuss several promising application scenarios of NOMA in next generation wireless networks. Specifically, we will discuss unmanned aerial vehicle (UAV) aided communications, robotic communications, and massive and critical machine-type communication (MC-MTC).

\subsection{NOMA-Based UAV-Aided Communications}
\begin{figure*}[t!]
\begin{center}
    \includegraphics[width=6in]{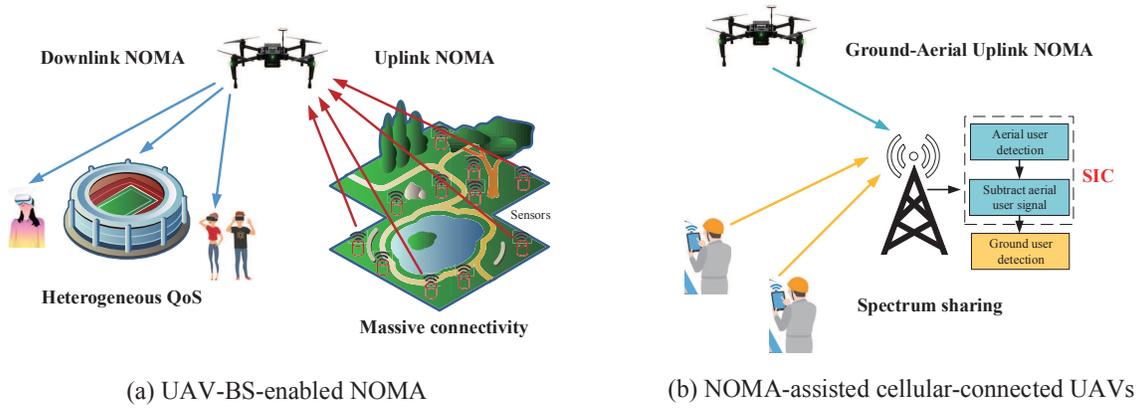}
    \caption{Two categories of NOMA-based UAV-aided communications.}
    \label{UAV application}
\end{center}
\end{figure*}
With the rapid development of manufacturing technology and the continuous reduction of costs, UAV-aided communications have been recognized as an emerging technique for next generation wireless networks due to the resulting high mobility and flexibility~\cite{yuanwei2019wcom,zhang2019tcom,8856258}. Equipped with communication devices, UAVs can act as BSs, relays, and users in wireless networks and facilitate a wide range of applications (e.g., wireless service recovery in emergencies, traffic offloading for temporary hotspots, environmental monitoring, and cargo delivery) by exploiting their unique line-of-sight (LoS) dominated air-to-ground (A2G) channel. However, these applications also present significant challenges, such as the management of the limited available wireless resources and the cancelation of the A2G interference. To cope with these issues, there are two promising approaches of employing NOMA in UAV-aided communications, namely, UAV-BS-enabled NOMA and NOMA-assisted cellular-connected UAVs, which are depicted in Fig.~\ref{UAV application}. A summary of the corresponding research contributions is provided in Table \ref{UAV-NOMA}.

\subsubsection{UAV-BS-enabled NOMA} As shown in Fig.~\ref{UAV application}(a), one common application scenario is that UAVs serve as aerial BSs to provide wireless service to areas where the terrestrial infrastructure is overloaded or even completely absent, such as outdoor temporary hotspots and data collection in the wild. In this category, employing NOMA can help UAV-BS to not only satisfy the corresponding heterogeneous QoS requirements since the users' channel conditions can be adjusted by changing the UAV's location, but can also facilitate massive connectivity since the encouragement of resource sharing in NOMA allows more users to be connected with the UAV-BS as compared to OMA. Motivated by these benefits, significant research efforts have been devoted to UAV-BS enabled NOMA. For instance, the authors of~\cite{houUAVNOMA} studied multi-UAV-enabled downlink NOMA in large-scale cellular networks, where two practical strategies were proposed, namely a user-centric strategy and a UAV-centric strategy. In the user-centric strategy, all ground users have to be served by the deployed UAVs, which can be regarded as an emergency communication use case (e.g., after disasters in remote areas or in rural areas). In the UAV-centric strategy, the UAVs aim to offload traffic from the terrestrial wireless networks (e.g., hotspots). For each strategy, exact expressions for the coverage probability were derived and numerical results confirmed that employing NOMA at the UAV-BSs can achieve a significant performance gain over OMA. Moreover, the setup was extended by considering multi-antenna UAV-BSs in \cite{houmultipleUAVNOMA}, where multiple ground users were served using a cluster-based NOMA design. Different from the quasi-static UAVs assumed in \cite{houUAVNOMA,houmultipleUAVNOMA}, the mobility of UAVs can be further exploited to improve the communication performance via UAV trajectory design. It is worth noting that this new DoF also introduces new challenges for designing UAV-BS enabled NOMA. This is because in NOMA transmission, the user decoding order is mainly determined by the users' channel qualities, which can dynamically change with the movement of the UAV-BS. This often leads to a highly-coupled joint UAV trajectory and NOMA decoding order optimization problem, which is non-trivial to solve. To this end, the authors of~\cite{cuiUAVNOMA} studied the joint UAV trajectory design and resource allocation problem, where a penalty dual-decomposition (PDD) technique was employed to deal with the downlink NOMA decoding order design. The obtained results showed that for the considered UAV-BS communication system, the performance gain of NOMA over OMA is more significant when the total UAV flight time is small. This underscores the effectiveness of employing NOMA in this use case since the maximum flight time of commercial UAVs is usually limited (e.g., around 30 minutes). As the limited on-board energy storage of UAVs is a major bottleneck in practical applications, the authors of~\cite{muUAVNOMA} further investigated the adoption of uplink NOMA in energy-constrained UAV data collection systems. In particular, a max-min UAV data collection rate problem was formulated by jointly optimizing the trajectory of the UAV and the scheduling and power control of the ground sensors, subject to the energy limitations of both the UAV and the ground sensors. The reported results also revealed that NOMA is more suitable for UAV data collection systems than OMA, especially when the UAV on-board energy storage is limited.
\subsubsection{NOMA-assisted cellular-connected UAVs} As shown in Fig.~\ref{UAV application}(b), UAVs can be integrated into existing wireless networks acting as aerial users, namely cellular-connected UAVs, and share the same spectrum with the ground users. Therefore, there is no dedicated spectrum resource allocated for cellular-connected UAVs, which is of vital importance for alleviating the spectrum shortage. However, cellular-connected UAVs will cause severe interference to the terrestrial wireless networks due to the spectrum sharing. To resolve this issue, a novel ground-aerial uplink NOMA framework was established in~\cite{Weidong2019jstsp,Xidong2020tcomuav}. As LoS dominated A2G channels are usually much stronger than the conventional terrestrial channels, the resulting asymmetric channel conditions allow the BS to first decode the aerial user's signal and subtract it from the received signal before decoding the ground users' signals, i.e., following the uplink NOMA protocol. Employing the ground-aerial uplink NOMA framework, the authors of~\cite{Weidong2019jstsp} proposed a uplink cooperative NOMA scheme to deal with the A2G interference from one static UAV user to a number of BSs, where some selected BSs first decode the UAV's signal and further deliver it to their neighboring BSs. The authors of~\cite{Xidong2020tcomuav} further investigated the trajectory design for cellular-connected UAVs, where the A2G interference can be canceled by the associated BS with the aid of NOMA and the UAV's trajectory was simultaneously optimized to control the A2G interference imposed on the non-associated BSs. To facilitate the resulting UAV trajectory design, two types of regions were defined, namely the uplink NOMA zone and the QoS protected zone. The uplink NOMA zone denotes a circular region, in which the UAV can be associated with the BS located in the center of this region. The QoS protected zone denotes a circular region, in which the UAV cannot enter if it is not associated with the BS located in the center. The authors of \cite{Xidong2020tcomuav} first proved that regardless of the amount of information bits that need to be delivered by the UAV to each BS, the optimal trajectory structure follows the fly-hover-fly protocol. Based on this insight, two algorithms were developed by employing graph theory and the SCA method. Furthermore, the ground-aerial uplink NOMA framework was extended to the case of multi-antenna UAVs in \cite{Pang_UAV_NOMA,19_ICC_Liu}, where the precoding vector of the UAV was optimized to maximize the achievable rate.
\begin{table*}[!t]\large
\caption{Summary of existing works on the application of NOMA in UAV-aided communications.}
\begin{center}
\centering
\resizebox{\textwidth}{!}{
\begin{tabular}{!{\vrule width1.5pt}l!{\vrule width1.5pt}l!{\vrule width1.5pt}l!{\vrule width1.5pt}l!{\vrule width1.5pt}}
\Xhline{1.5pt}
\centering
\makecell[c]{\textbf{Category}}  & \makecell[c]{\textbf{Reference}} &\makecell[c]{\textbf{UAV Setup}} & \makecell[c]{\textbf{Characteristic/Technique}} \\
\Xhline{1.5pt}
\centering
\multirow{4}{*}{\makecell[c]{UAV-BS enabled NOMA}} & \makecell[c]{\cite{houUAVNOMA}} & {Multiple static single-antenna UAVs} & {Two strategies for UAV providing wireless service}  \\
\cline{2-4}
\centering
& \makecell[c]{\cite{houmultipleUAVNOMA}}  &{One static multi-antenna UAV}  &{Cluster-based NOMA for A2G communications} \\
\cline{2-4}
\centering
& \makecell[c]{\cite{cuiUAVNOMA}}   & {One mobile single-antenna UAV}    & {PDD based algorithm for joint UAV trajectory and NOMA decoding order design} \\
\cline{2-4}
\centering
& \makecell[c]{\cite{muUAVNOMA}}   & {One mobile single-antenna UAV}    & {Double energy limitations at both the UAV and ground sensors} \\
\Xhline{1.5pt}
\centering
\multirow{3}{*}{\makecell[c]{NOMA assisted cellular-connected UAV}}& \makecell[c]{\cite{Weidong2019jstsp}}   & {One static single-antenna UAV}    & {Uplink cooperative NOMA for A2G interference cancelation} \\
\cline{2-4}
\centering
& \makecell[c]{\cite{Xidong2020tcomuav}}   & {One mobile single-antenna UAV}    & {Interference-aware UAV trajectory design} \\
\cline{2-4}
\centering
& \makecell[c]{\cite{Pang_UAV_NOMA,19_ICC_Liu}}   & {One static multi-antenna UAV}    & {Uplink precoding design for A2G interference cancelation} \\
\Xhline{1.5pt}
\end{tabular}
}
\end{center}
\label{UAV-NOMA}
\end{table*}
\subsection{NOMA-Enhanced Robotic Communications}
\begin{figure}[t!]
\centering
\subfigure[OMA]{\label{OMA}
\includegraphics[width= 2.6in]{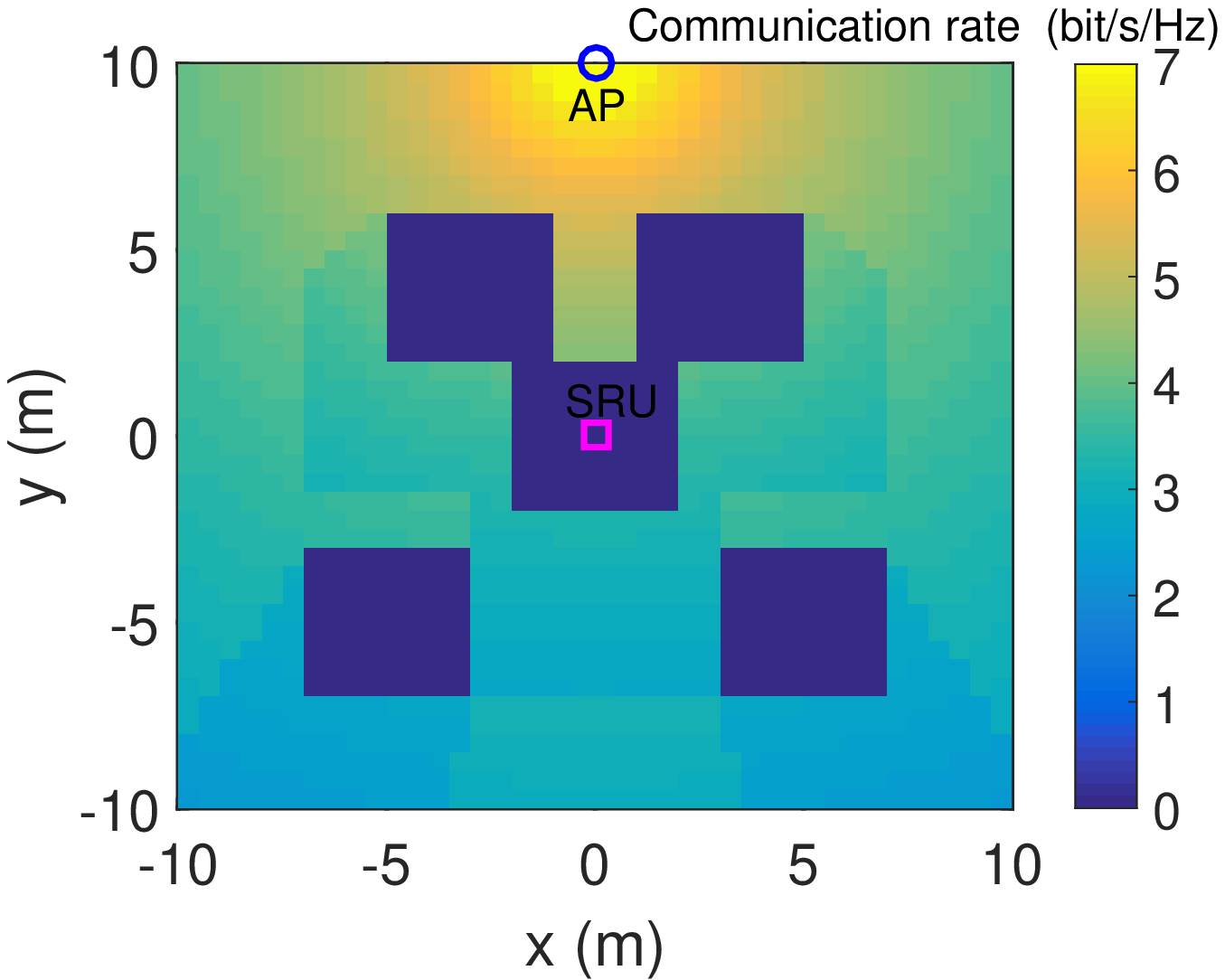}}
\subfigure[NOMA]{\label{NOMA}
\includegraphics[width= 2.6in]{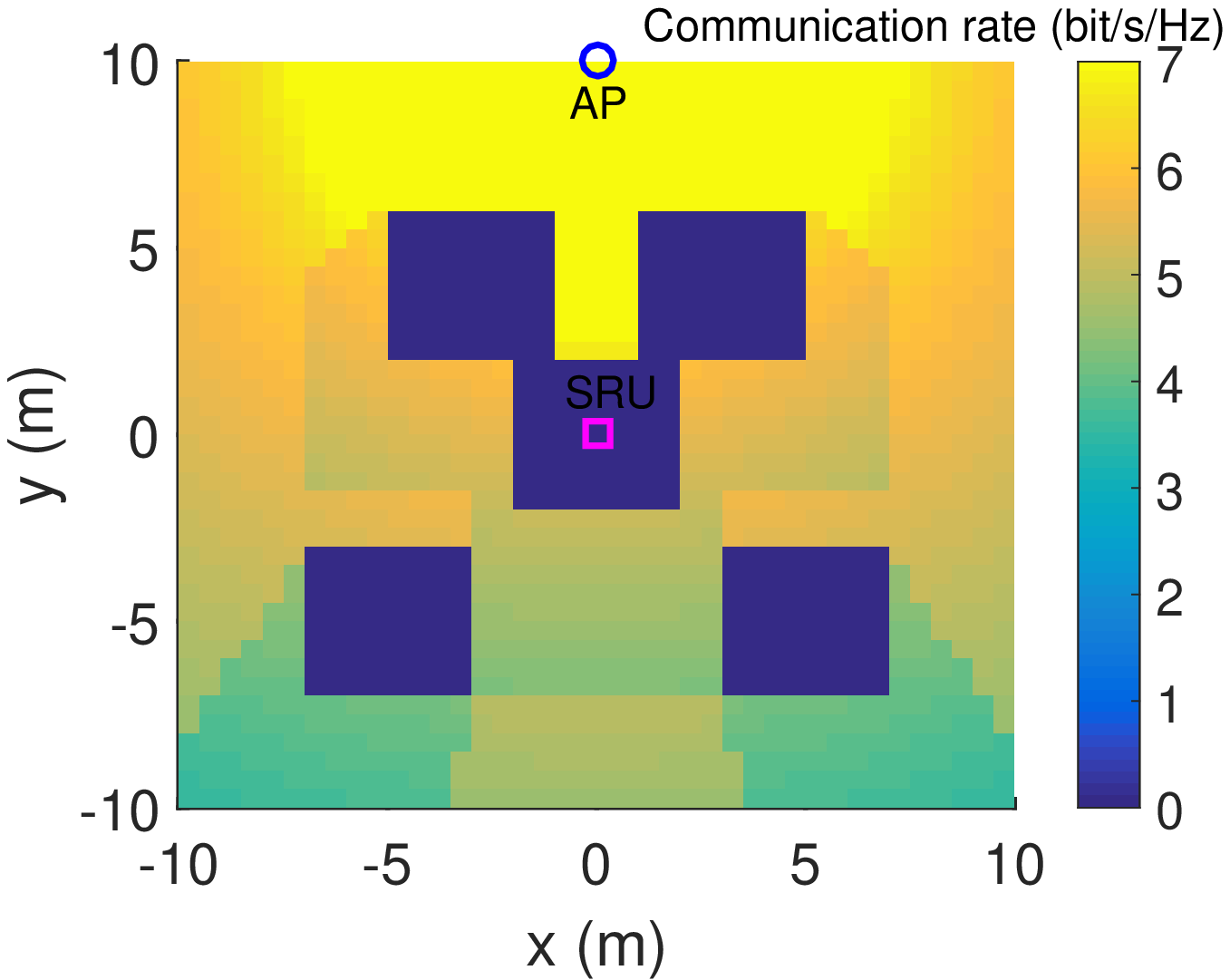}}
\setlength{\abovecaptionskip}{-0cm}
\caption{Illustration of the obtained communication rate maps for OMA and NOMA, where the AP and the static user are located at $\left( {0,10,2} \right)$ meter and $\left( {0,0,1.3} \right)$ meter, respectively. The five dark blue regions are covered by obstacles with a height of 1.3 meter. The other parameters adopted can be found in~\cite[Section VI]{xidong_robot}.}\label{rate map}
\end{figure}

With the rapid development of computer science and robotic techniques, robots will play an increasingly important role in human society in different fields such as smart home and smart factory. Among others, one promising direction is to integrate robots into wireless networks as robotic users, namely connected robots~\cite{robot}. By doing so, connected robots can accomplish missions relying on information exchange with operators (e.g., Access Points (APs) and BSs) instead of relying on their own computational resources, thus being more cost-efficient and less computation-constrained. Despite following a similar concept as cellular-connected UAVs, the application scenarios of connected robots are more challenging. On the one hand, the communication channels of connected robots are more likely to be blocked than the LoS dominated A2G channels, thus leading to rapidly changing channel conditions. On the other hand, the number of connected robots in the network is usually large, and each of them may have different working conditions (e.g., static and mobile) and communication requirements (e.g., data-hungry or delay-insensitive). The resulting heterogeneous communication scenario makes the resource management a non-trivial task. To cope with the aforementioned issues, some initial research efforts have begun to exploit NOMA for robotic communication~\cite{xidong_robot,xinyu_robot}. In particular, the authors of \cite{xidong_robot} investigated an indoor communication-aware robot path planning problem, where one mobile robotic user (MRU) and one static robotic user (SRU) are simultaneously served by an AP via NOMA. The MRU is dispatched to travel from a predefined initial location to the final location, subject to a required communication quality. To minimize the required travelling time, a radio map based approach was proposed to characterize the location-dependent communication quality in the considered indoor environment. Fig. \ref{rate map} depicts the communication rate maps obtained for NOMA and OMA, which characterize the spatial distribution of the expected rate of the MRU while satisfying the communication rate requirement of the SRU. As can be seen from Fig. \ref{OMA}, with OMA, only a small region can achieve a rate of more than 5 bit/s/Hz for the MRU. However, in Fig. \ref{NOMA}, it can be observed that more than half of the region can achieve a rate of more than 5 bit/s/Hz if NOMA is employed. Therefore, a significant rate gain can be achieved by NOMA over OMA. The considered setup was extended to path planning for multiple mobile robots in \cite{xinyu_robot}.
\begin{figure*}[t!]
\begin{center}
    \includegraphics[width=7in]{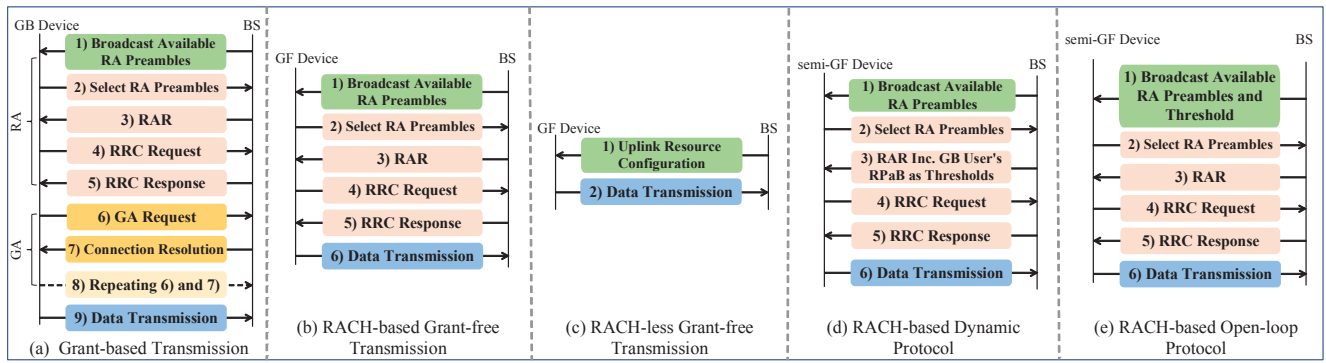}
    \caption{A comparison of GF, GB, and semi-GF transmission schemes.}
    \label{SGF}
\end{center}
\end{figure*}
\subsection{Massive and Critical Machine-Type Communication (MC-MTC)}
Most 6G services including several vertical sectors, e.g., intelligent transport systems, industry 4.0, and smart cities, are gradually being digitalized. MTC is one of the most promising enablers for supporting this large-scale digitization. In addition to the traditional ``massive'' demand of MTC, the strict QoS requirement becomes another key performance indicator (KPI) for MTC in 6G, which evolves the current MTC towards MC-MTC~\cite{mahmood2020white}. These two KPIs cause new challenges for 6G, especially for uplink transmission with finite blocklength, where ideal error-free decoding is not possible, as discussed in Section III.B.2. One possible solution for reducing the decoding error probability and satisfying the massive and critical requirements in MC-MTC is to develop advanced random access techniques. By abandoning scheduling requests and grants, which may cause a large signaling overhead, Grant-Free (GF) random access transmission is able to enhance connectivity compared to conventional Grant-Based (GB) transmission. As shown in Figs. \ref{SGF}(a)-(c), the main difference between GF and GB transmission is that GF transmission does not employ a grant acquisition process. Thus, with GF transmission, devices can transmit messages without waiting for permission. More specifically, GF transmission can be divided into two categories based on whether a random access process is employed or not, namely, Random Access Channel (RACH)-based GF transmission and RACH-less GF transmission. However, when the required data rate for the GF users is low or the channel conditions of GF users are weak, the SE of GF transmission becomes extremely limited. Fortunately, NOMA is able to handle this problem by adjusting the power resources to balance the traffic of users with different channel conditions~\cite{8443376}. Besides GF-NOMA, another random access NOMA scheme, namely semi-GF NOMA, has been proposed in~\cite{8662677}. Note that for GF transmission, the QoS requirements of the users can not be guaranteed. To address this issue, in semi-GF NOMA, GB users and GF users are grouped into one NOMA cluster by allocating the spare spectrum resources of GB users to GF users. More specifically, a threshold is introduced for semi-GF transmission to determine which portion of the users is to employ semi-GF transmission, while the remaining users employ GB transmission. In Figs. \ref{SGF}(d) and (e), two semi-GF protocols are presented based on RACH-based GF transmission, namely RACH-based dynamic protocol and RACH-based open-loop protocol utilizing different access thresholds, i.e., the instantaneous Received Power at the BS (RPaB) and an average threshold~\cite{9244136}. These schemes can not only ensure the success of GB users' transmissions but can also provide additional access channels for GF users at the same time, thus satisfying both the ``massive'' and ``critical'' KPIs of MC-MTC.\\
\indent There are several successful examples of random access based NOMA, including GF-NOMA and Semi-GF NOMA schemes, for MTC. Since traditional uplink NOMA employs a fixed transmit power, collisions are severe for GF-NOMA. To overcome this drawback, the authors of~\cite{yi2020multiple} introduced a multi-transmit-power framework for random access NOMA. The transmit power levels, which form a power pool, are connected to the geographic locations of the users due to the severe impact of path loss on the channel quality in wireless communications. The obtained results showed that by offering multiple transmit power levels to users, the degradation caused by collisions can be significantly reduced. Moreover, the authors of~\cite{9452792} designed an open-loop power control for the power pool with the aid of Deep Q-Network (DQN), where the power pool was first trained offline and then updated online to control the performance-complexity tradeoff. The proposed GF-NOMA system outperforms networks with OMA in terms of the achievable data rates. For semi-GF NOMA, the authors of~\cite{ding2020a} proposed a QoS-guarantee method to enhance the stability of semi-GF transmissions. The ergodic rate performance of semi-GF NOMA systems was evaluated in~\cite{9244136}. The authors of~\cite{fayaz2021competitive} further extended the power pool concept to semi-GF NOMA systems by considering the competitive nature of IoT devices. In this work, a competitive multi-agent DQN was exploited for distributed power control. The proposed semi-GF NOMA IoT system performs better than systems with fixed power control and networks with pure GF transmission. However, current works mainly focus on throughput performance. For other MTC QoS metrics, e.g., latency, reliability, etc., efficient random access NOMA concepts have yet to be developed.
\subsection{Other Application Scenarios}
Besides the above discussed application scenarios of NOMA in future wireless networks, there are many other interesting 6G scenarios, in which NOMA can be fruitfully employed. We focus on two topics in the following.
\subsubsection{Mobile Edge Computing (MEC) Networks}
The explosive growth of mobile applications and services in next generation wireless networks requires unprecedentedly high data transmission speed and low latency. As a result, MEC has emerged as a possible solution, which moves computing resources from the central network towards the network edges, thus relieving the backhaul burden on APs~\cite{mao2017ICST}. Compared to conventional MEC networks relying on OMA, NOMA provides benefits for both computational task transmission and computation result announcement. In particular, mobile devices can simultaneously offload their computation tasks to the AP using the uplink NOMA protocol, while the computation results can be downloaded from the AP to mobile devices exploiting the downlink NOMA protocol. Due to the resource sharing characteristic of NOMA, employing NOMA in MEC networks enables highly flexible resource allocation, thus significantly improving the computation performance. Motivated by these advantages, the authors of \cite{Ding2019tcom} carried out various asymptotic performance analyses to evaluate the impact of the channel conditions and transmit powers on the performance of NOMA-MEC networks. To obtain a joint communication and computation resource allocation solution for NOMA-MEC networks, the authors of~\cite{Wang2019tcom} and~\cite{song2018comletter} minimized the energy consumption of all users subject to a constraint on the task execution latency in heterogeneous networks, where the Lagrange dual method and an iterative algorithm were applied, respectively. The overall computation delay minimization problem in a NOMA-enabled MEC network was studied in \cite{8611381}, where the users' offloaded workloads, the task uploading durations, and the computation-results downloading durations were jointly optimized.
\subsubsection{E-health}
Nowadays, human health problems are steadily increasing medical requirements, which calls for highly efficient communication networks to facilitate the connection between patients and hospitals. Relevant use cases include real-time tele-medicine/tele-surgery for emergencies and rural areas~\cite{liu2021application,Di20195Ghealth}. The application of NOMA in e-health is appealing for coordination of the resource allocation among numerous smart devices (e.g., smart watches and sensors), which is indispensable for establishment of efficient monitoring systems for e-health. To facilitate this design, the authors of~\cite{Xuewan2021jsacc} proposed a multi-carrier NOMA scheme for in-home health networks, which allows more monitoring devices to be connected and more information bits to be received compared to OMA based designs.
\section{Interplay Between NOMA and Other Emerging Physical Layer Techniques}
Compatibility is one of the most important assessment criteria for a new technology. Since 6G will combine various different physical layer technologies, it is essential and challenging to investigate the interplay between NOMA and other emerging physical layer techniques. This section mainly focuses on the integration of NOMA with other techniques and surveys existing works in these areas.
\subsection{Reconfigurable Intelligent Surface (RIS)-NOMA}
\begin{figure}[t!]
\begin{center}
    \includegraphics[width=3in]{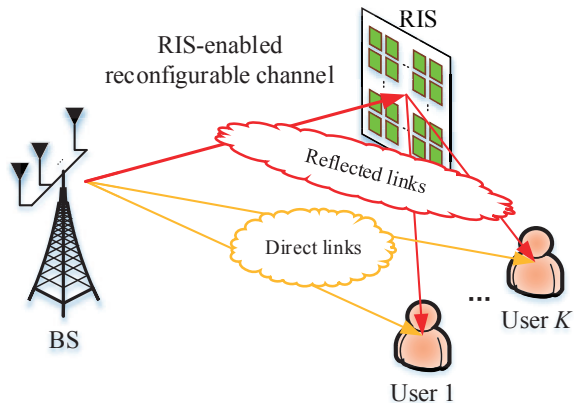}
    \caption{Illustration of RIS-enabled wireless communications.}
    \label{RIS}
\end{center}
\end{figure}
\begin{figure}[t!]
\begin{center}
    \includegraphics[width=3in]{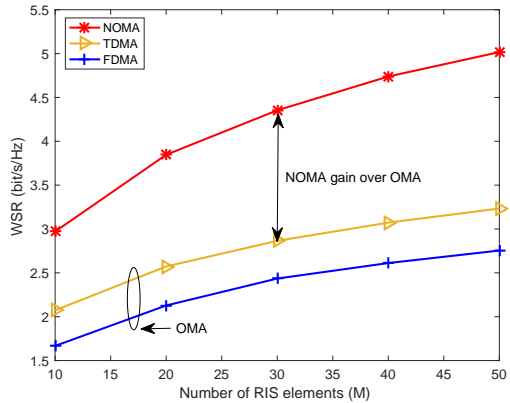}
    \caption{Weighted sum rate versus the number of RIS elements in an RIS-aided 4-user communication system. The AP is located at ${\left( {{0},{0},{5}} \right)}$ meter and the $k$th user is located at ${\left( {{25+5k},{0},{1.5}} \right)}$ meter with rate weights of ${w_k} = 0.1k$, $\forall k \in \left\{ {1,2,3,4} \right\}$. The location of the RIS is optimized along $\left\{ {\left( {x,5,5} \right)|30 \le x \le 45} \right\}$ meter. The direct links between the AP and the users are assumed to be blocked. The other parameters adopted can be found in~\cite[Section VI]{Mu_deployment}.}.
    \label{WSR}
\end{center}
\end{figure}

Motivated by the development of metasurfaces, RISs~\cite{RIS_survey}, also known as intelligent reflecting surfaces (IRSs)~\cite{IRS_survey}, have recently emerged as a promising technique for the next generation wireless network. An RIS is a man-made thin surface, which consists of a large number of low-cost reconfigurable elements (e.g., positive intrinsic negative (PIN) diodes) and a smart controller (e.g., a field programmable gate array (FPGA)). By configuring the phase and possibly the amplitude responses of the RIS elements, the propagation of the incident wireless signal can be beneficially modified, thus creating a smart radio environment~\cite{Renzo_IRS}, as shown in Fig. \ref{RIS}. Compared to conventional multi-antenna techniques and active relays, the hardware cost and power consumption of RISs is significantly lower since RISs do not require RF chains. Moreover, RISs can be easily integrated into the existing wireless networks. Structures on which RISs can be deployed include but are not limited to building facades, indoor walls, roadside billboards, and windows. Given the above favorable properties of RISs, extensive research efforts have been devoted to investigating the benefits of RIS-aided communication systems, such as the mitigation of interference~\cite{Dongfang}, the reduction of transmit power~\cite{Wu_IRS}, and the enhancement of communication secrecy~\cite{Yu_secure}. Among others, the combination of RISs and NOMA is an appealing option for the following two reasons~\cite{IRS-NOMA}. First, recall the fact that the performance gain of NOMA over OMA mainly depends on the distinct channel conditions of the involved users, which may not always hold in practice. By carefully selecting the deployment location and reflection coefficients of RISs, the channel disparity among the users can be enlarged instead of being solely determined by the wireless environment, thus yielding a higher NOMA gain. Second, the ``channel changing'' characteristic provided by RISs can facilitate a smart QoS-based NOMA operation. To be more specific, in conventional NOMA communication systems without RISs, the NOMA decoding order for users mainly depends on the order of their given channel conditions, which generally cannot be influenced. Therefore, only users with higher NOMA decoding orders can achieve high communication rates by using SIC to cancel inter-user interference, while users with lower NOMA decoding orders only achieve low data rates due to the unmitigated inter-user interference. However, in practice, the QoS requirements of the users may not be consistent with the order of their channel conditions. In other words, a user with a weaker channel may require a higher data rate than a user with a stronger channel. RISs can help to overcome this issue. In particular, we can order the users in terms of their QoS requirements and RISs can be employed to realize the corresponding desired order of channel conditions, i.e., facilitating RIS-aided QoS-based NOMA communication.\\
\begin{table*}[!t]\large
\caption{Summary of existing research contributions on RIS-NOMA.}
\begin{center}
\centering
\resizebox{\textwidth}{!}{
\begin{tabular}{!{\vrule width1.5pt}l!{\vrule width1.5pt}l!{\vrule width1.5pt}l!{\vrule width1.5pt}l!{\vrule width1.5pt}}
\Xhline{1.5pt}
\centering
\makecell[c]{\textbf{Reference}}  & \makecell[c]{\textbf{Setup}} &\makecell[c]{\textbf{Main Objective}} & \makecell[c]{\textbf{Characteristic/Technique}} \\
\Xhline{1.5pt}
\centering
\makecell[c]{\cite{Mu_capacity}} & \makecell[c]{SISO} & {Capacity region} & {Dynamic RIS configuration with NOMA is the capacity-achieving transmission strategy}  \\
\hline
\centering
\makecell[c]{\cite{shuowen_capacity}}& \makecell[c]{SISO}  &{Capacity region}  &{Rate-profile and uplink-downlink duality based capacity region characterization for RIS-aided MAC and BC} \\
\hline
\centering
\makecell[c]{\cite{Mu_deployment}}& \makecell[c]{SISO}   & {Weighted sum rate}    & {Asymmetric and symmetric deployment strategies are preferred by NOMA and OMA, respectively} \\
\hline
\centering
\makecell[c]{\cite{Mu_IRS_NOMA}}& \makecell[c]{MISO}   & {Sum rate}    & {Three types of RIS elements based on the configuration of amplitude and phase shift coefficients} \\
\hline
\centering
\makecell[c]{\cite{yang_RIS_NOMA}/\cite{fu_RIS_NOMA}}& \makecell[c]{MISO}   & {User fairness/Transmit Power}    & {Low-complexity and near-optimal user ordering scheme} \\
\hline
\centering
\makecell[c]{\cite{zhu_RIS_NOMA}}& \makecell[c]{MISO}   & {Transmit power}    & {Extended quasi-degradation condition} \\
\hline
\centering
\makecell[c]{\cite{Hou_RIS_NOMA}}& \makecell[c]{MIMO}   & {Outage probability analysis}    & {Signal cancelation based RIS phase shift design} \\
\hline
\centering
\makecell[c]{\cite{Ding_RIS_NOMA}}& \makecell[c]{MISO}   & {Outage probability analysis}    & {Multiple RISs for cell-edge users to extend coverage} \\
\hline
\centering
\makecell[c]{\cite{Mu_UAV_RIS_NOMA}}&\makecell[c]{Multi-cell SISO}   & {Sum rate}    & {Simultaneous intra-cell signal enhancement and inter-cell interference mitigation for A2G communications} \\
\Xhline{1.5pt}
\end{tabular}
}
\end{center}
\label{RIS-NOMA}
\end{table*}
\indent The aforementioned advantages of RIS-NOMA have drawn significant research interests in an effort to exploit the new DoFs introduced to NOMA by RISs. A summary of existing research contributions on RIS-NOMA is provided in Table \ref{RIS-NOMA}. In particular, the authors of \cite{Mu_capacity} considered an RIS-aided multi-user SISO BC. Different from current RIS work assume the static RIS configuration (i.e., the RIS reflection coefficients can only be adjusted once during one transmission), the dynamic RIS configuration is employed, where the RIS reflection coefficients can be adjusted multiple times during one transmission. The capacity and rate regions achieved by NOMA and OMA were characterized by jointly optimizing the power allocation of the AP and the phase shift of the RIS in each time slot. The derived solutions revealed that the optimal transmission strategy for OMA is alternating transmission between all users with their effective channel power gains maximized by the RIS. For NOMA, the optimal transmission strategy is alternating transmission between different user groups with dynamic decoding orders facilitated by the RIS. Moreover, the results in \cite{Mu_capacity} also showed that the capacity gain of NOMA is less sensitive to the dynamic RIS configuration than the rate gain of OMA. This revealed that, compared to OMA, NOMA not only achieves a higher capacity gain but also requires less hardware complexity at the RIS, because the dynamic RIS configuration needed for OMA requires considerable overhead for information exchange between the AP and the RIS. The capacity/rate region characterization problem of the RIS-aided two-user SISO MAC and BC with NOMA and OMA was further investigated in \cite{shuowen_capacity} by leveraging the rate-profile method and uplink-downlink duality, where two types of network-level RIS deployment strategies were considered, namely centralized deployment and distributed deployment. In particular, one large-size RIS was deployed near the AP for the centralized deployment strategy, while two small-size RISs were separately deployed near each user for the distributed deployment strategy. In both strategies, the total number of RIS elements was assumed to be the same for a fair comparison. One interesting insight obtained in \cite{shuowen_capacity} is that regardless of the employed multiple access scheme, the centralized deployment strategy always outperforms the distributed one in terms of the capacity/rate region. As a further advancement, the authors of \cite{Mu_deployment} investigated the link-level optimal RIS deployment design for an RIS-aided multi-user downlink network. Considering NOMA, TDMA, and FDMA transmission, a weighted sum rate maximization problem was formulated by jointly optimizing the power allocation of the AP and the phase shifts and deployment location of the RIS. Both monotonic optimization and alternating optimization based algorithms were developed to find a performance upper bound and a suboptimal solution, respectively, which were shown to achieve a similar performance via simulations. One useful guideline for RIS deployment revealed in \cite{Mu_deployment} was that an asymmetric RIS deployment strategy is preferable for NOMA, while a symmetric RIS deployment strategy is superior for OMA. In Fig. \ref{WSR}, we present the maximum weighted sum rate obtained by different multiple access schemes in \cite{Mu_deployment}. As can be observed, NOMA significantly outperforms both TDMA and FDMA. The performance gain is more pronounced as the number of RIS elements increases, which confirms the advantages of the integration of NOMA and RISs. Moreover, the authors of \cite{chenyu} studied simultaneously transmitting and reflecting RIS (STAR-RIS) assisted communication systems. Compared to reflecting-only RIS, the signal incident on STAR-RIS can be transmitted and reflected to both sides~\cite{STAR_mag,STAR_xidong}. By exploiting this unique feature, the fundamental transmission-versus-reflection coverage tradeoff was characterized for NOMA and OMA in \cite{chenyu}. The obtained results showed that NOMA yields a more significant gain over OMA for STAR-RIS than for conventional RIS.\\
\indent Furthermore, the investigation of RIS-NOMA was extended to multi-antenna transmission, which additionally requires the design of an active beamformer. For instance, the authors of \cite{Mu_IRS_NOMA} studied an RIS-aided MISO downlink NOMA network, where the joint active and passive beamformer design problem was formulated for the maximization of the sum rate of all users. In particular, three types of RIS elements were considered, which differed in whether the amplitude and phase shift coefficients can be continuously adjusted or not. Since the RIS can reconfigure the users' channels, the optimal NOMA decoding order in \cite{Mu_IRS_NOMA} was obtained via exhaustive search, which may lead to an unacceptable computational complexity. To address this issue, some low-complexity user ordering schemes were proposed in \cite{yang_RIS_NOMA,fu_RIS_NOMA}, and were shown to be able to achieve a similar performance as the exhaustive search. Inspired by the observation that for multi-user MISO transmission, NOMA can achieve the same performance as DPC when the quasi-degradation condition is satisfied~\cite{quasi-degradation}, the authors of \cite{zhu_RIS_NOMA} exploited this condition in RIS-aided communication systems. Their results revealed that the quasi-degradation condition can be satisfied with a high probability by optimizing the phase shifts of the RIS, thus ensuring that NOMA can achieve high performance. As a further step, some novel transmission frameworks were proposed in~\cite{Hou_RIS_NOMA,Ding_RIS_NOMA,Mu_UAV_RIS_NOMA}. In~\cite{Hou_RIS_NOMA}, a signal cancellation-based design was proposed for RIS-aided MIMO-NOMA networks, where the reflection coefficients of the RIS were designed to mitigate the inter-cluster interference received by different NOMA clusters. The results revealed that, owing to the additional signal cancelation capability enabled by the RIS, the constraints regarding the number of antennas at the transmitters and the number of antennas at the receivers can be relaxed, which, however, have to be strictly satisfied for conventional MIMO-NOMA networks. We note that the coverage of one RIS is limited. Hence, one RIS may not be able to cover all users in the network. To overcome this issue, in~\cite{Ding_RIS_NOMA}, multiple distributed RISs were deployed to establish a reflection link between the BS and cell-edge users, each of which shared the same beamformer as one cell-center user to facilitate NOMA. By doing so, additional cell-edge users can be served in an ``add-on'' manner. The authors of \cite{Mu_UAV_RIS_NOMA} further established a novel RIS-enhanced multi-UAV NOMA transmission framework, where the RIS was deployed to enhance the strength of the desired signal between the UAV and its intended users while mitigating the strong LoS dominated interference between the UAV and unintended users. Simulation results showed that significant sum rate gains can be achieved by NOMA over OMA.
\subsection{Orthogonal Time Frequency Space (OTFS)-NOMA}
\begin{figure}[t!]
\begin{center}
    \includegraphics[width=3in]{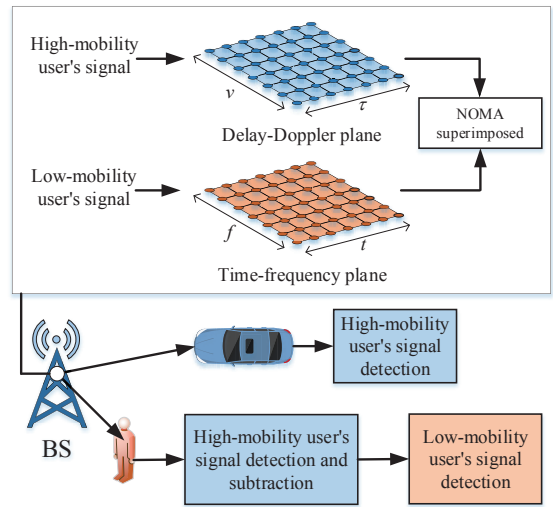}
    \caption{Illustration of a downlink OTFS-NOMA system with one high-mobility user and one low-mobility user.}
    \label{OTFS}
\end{center}
\end{figure}

Supporting users with different mobility profiles (e.g., UAVs, cars, and high-speed trains) is an important feature of next generation wireless networks. However, supporting these users in a spectrally efficient manner is difficult to accomplish. Take a user with high mobility as an example, where the user's high mobility introduces two challenging issues. One issue are Doppler frequency shifts, which are particularly damaging in current 4G and 5G systems, because Doppler frequency shifts cause inter-carrier interference for OFDM modulated signals and can severely degrade the system performance. The other issue is how to realize reliable and timely channel estimation, where a huge amount of system overhead might be wasted as pilot signals have to be transmitted frequently. Recently, a new approach, termed orthogonal time frequency space (OTFS) modulation, has been proposed to support high-mobility users~\cite{hadani2018otfs,8516353,8503182}. The key idea is to place the high-mobility users' signals in the delay-Doppler plane. Thereby, the users' channels, which are time varying in the time-frequency plane, are converted to time invariant channels in the delay-Doppler plane. As a result, channel estimation can be straightforwardly carried out in the delay-Doppler plane.\\
\indent However, the performance of OTFS is limited by the OTFS resolution, which measures how accurately a user's channel can be located in the delay-Doppler plane. In general, the OTFS resolution can be improved by asking a user to transmit for a long time duration and over many different frequency channels. In other words, the use of OTFS can result in the situation that the bandwidth resources and time slots are occupied by high-mobility users. If these users have poor channel conditions or do not need to be served at high data rates, the SE of OTFS can be low. To address this issue, the authors of~\cite{8786203} proposed a novel OTFS-NOMA scheme to improve the SE for the scenario with heterogenous user mobilities, where the high-mobility users' signals are placed in the delay-Doppler plane and the low-mobility users' signals are placed in the time-frequency plane. The corresponding two-user downlink OTFS-NOMA system is depicted in Fig. \ref{OTFS}. With this approach, the high-mobility users can enjoy the benefit of time-invariant channels in the delay-Doppler plane, and the low-mobility users are additionally served by placing their signals into the time-frequency plane, where the interference caused by the signals in the two non-orthogonal planes can be tackled by applying the well-known NOMA interference management techniques. For MIMO systems, the authors of~\cite{8901184} further designed a robust and low-complexity beamforming approach for OTFS-NOMA systems, which aims to guarantee the QoS of the high-mobility users while maximizing the data rate of the low-mobility users. The proposed beamforming approach provides a significant performance gain over random beamforming. Since the research on OTFS-NOMA systems is still in its infancy, there are numerous open problems, which require further research efforts, e.g., joint optimization of power allocation and beamforming, user clustering based on the mobility profiles of the users, user handovers across different NOMA clusters.
\subsection{Integrated Sensing and Communication (ISaC)-NOMA}
Integrated sensing and communication (ISaC) has recently been recognized as an important technique for 6G since it provides new sensing capabilities and reduces the weight of the equipment in current wireless networks~\cite{8999605,ISAC}. These advantages are suitable for several mobile 6G scenarios: vehicle-to-everything (V2X), UAV, etc. Current ISaC systems include three main types: Communication-centric designs (CCD), sensing-centric designs (RCD), and joint designs (JD)~\cite{zhang2021overview}. In CCD, communication has the highest priority and radar sensing is an add-on function, which can tolerate poor detection performance. In SCD, the opposite applies. In JD, both the communication and sensing functions have the same priority, and hence controllable performance trade-offs are indispensable. Since sensing and communication signals employ different waveforms, their combination in the same time/frequency resource block is a challenging task. Moreover, when receiving the superposed radar echoes and communication signals in the same time/frequency resource block, it is difficult to simultaneously decode the communication signals and analyze the radar echoes. These two problems are obstacles to the development of ISaC. Fortunately, NOMA is able to offer additional access channels in a given time/frequency resource block, which potentially solves these problems. For example, two different waveforms can be labelled with different power levels. Iterative SIC or parallel SIC techniques can help to separate communication and radar signals at the receivers. From the perspective of NOMA, ISaC introduces a new mechanism of acquiring CSI, which improves the quality of NOMA transmission. For NOMA-enabled ISaC, the related research is in a very early stage. The authors of~\cite{9295417} explored the possibility of integrating different waveforms into one NOMA cluster. The proposed solution outperforms conventional OFDM-ISaC. Another idea for NOMA-ISaC systems is to use NOMA for controlling the signal and radar power, such that the prediction accuracy requirement can be relaxed as discussed in~\cite{7279172}. As radar signals are not modulated, the first SIC stage of ISaC-NOMA systems decodes the communication signals and the second SIC stage analyzes the remaining radar echoes~\cite{liu2021application} with the communication signal subtracted. The authors of \cite{ISAC_xidong} further considered NOMA-aided joint radar and multicast-unicast communication systems, where the BS transmits the mixed multicast-unicast information to both the radar-centric and communication-centric users employing NOMA while the superimposed signal is used to detect the radar-centric user. Numerical results showed that the proposed NOMA scheme significantly outperforms conventional SDMA and TDMA schemes in terms of both radar detection and communication performance.

\subsection{Other Potential Interplays}

\subsubsection{Coordinated Multi-Point (CoMP)-NOMA} CoMP transmission is an effective technique to mitigate inter-cell-interference (ICI) and improve the performance of cell-edge users~\cite{8760269}. Integrating NOMA with CoMP has two main benefits. On the one hand, for multi-cell NOMA communication systems, cell-edge NOMA users suffer from not only intra-cell NOMA interference but also inter-cell interference, which causes performance degradations. To this end, CoMP can help to cancel or partially cancel ICI for cell-edge NOMA users to improve communication performance. On the other hand, NOMA provides additional flexibility for CoMP to improve the performance of cell-centre users by allowing resource sharing between the cell-centre and cell-edge users and between different cells. To explore these benefits, the power allocation problem in CoMP-NOMA networks was studied by the authors of~\cite{8352643} to maximize the sum rate. To reduce complexity, this work proposed a distributed optimization method, where each BS used the Karush-Kuhn-Tucker approach to optimize its power allocation without the need to know the transmission strategies of the other coordinated BSs. For CoMP-NOMA with joint transmission, the authors of~\cite{8352618} proposed three novel user pairing strategies to maximize the network throughput. Furthermore, the authors of~\cite{8781867} proposed a generalized CoMP-NOMA transmission framework, where joint transmission can be applied for both cell-centre and cell-edge users.
\subsubsection{Full-Duplex (FD)-NOMA} FD is an important technique for next generation wireless networks, which allows simultaneous downlink and uplink transmission over the same time/frequency resources. The integration of FD with NOMA is promising to further improve the resource efficiency by allowing a large number of users to simultaneously download and/or upload information from and/or to the BS. Despite these appealing advantages, FD-NOMA also introduces challenges for interference management \cite{17_TCOM_Sun}, including the self-interference at the FD transmitter and the inter-user interference between the downlink and uplink NOMA users. Therefore, sophisticated resource allocation schemes have to be developed. For instance, the authors of \cite{17_TCOM_Sun} studied an FD multi-carrier NOMA system, where the power allocation and the subcarrier assignment were jointly optimized to maximize the weighted system throughput. To confirm the effectiveness of FD-NOMA, the authors of \cite{8306094} analyzed the outage performance of FD-NOMA and half-duplex (HD) NOMA. Numerical results showed that FD-NOMA can offer significant performance gains over HD-NOMA and OMA as long as the co-channel interference can be sufficiently suppressed, which underscores the importance of interference mitigation in FD-NOMA. Moreover, the authors of \cite{FD1} studied HD relay assisted NOMA networks, where two relay selection schemes were proposed and the corresponding outage performance was analyzed. Based on cooperative NOMA transmission, the authors of \cite{FD2} proposed to make the near user serve as a HD/FD relay to assist the far user. It was shown that FD NOMA is superior to HD NOMA for low SNR region.
\subsubsection{Visible Light Communications (VLC)-NOMA} For indoor environments, VLC is promising because of its usage of unlicensed spectrum, efficient implementation, high confidentiality, and low energy consumption~\cite{7792590}. Practical VLC has three main features: 1) The number of users served by each transmitter is limited; 2) the corresponding transmit SNR can be high; and 3) the channels in VLC are usually constant and can be easily estimated by the transmitter with high accuracy. However, one main drawback of VLC is that the modulation bandwidth is small. Therefore, multiple access techniques are indispensable for VLC to enhance the SE. On the one hand, from the perspective of VLC, NOMA is an attractive multiple access technique due to the following reasons. First, the limited number of users leads to a small number of SIC stages for NOMA, and hence the outage probabilities for all SIC stages can be easily controlled to an ultra-low level. Second, the high transmit SNR of VLC enables high NOMA-introduced throughput gains. Third, the slow-changing CSI of the users can be estimated with low overhead at the transmitter side, which reduces the complexity required for NOMA transmission. On the other hand, from the perspective of NOMA, VLC is also promising since it introduces two novel controllable parameters, namely, the transmission angles and the field of view, which can be exploited to achieve significant channel differences. To explore the interplay between NOMA and VLC, the authors of \cite{7572968} proposed a mathematical framework for evaluation of the average performance of NOMA-VLC systems. It was analytically shown that NOMA-VLC outperforms OMA-VLC in terms of the ergodic sum rate, where the gain are especially pronounced in the high SNR regime. By enlarging the channel gain difference between the paired users and choosing suitable light emitting diodes, the NOMA-induced gain can be further enhanced. For MIMO-VLC systems, the authors of \cite{8233180} proposed a normalized gain difference power allocation (NGDPA) method to increase the achievable sum rate. This work illustrated the superiority of employing NOMA in MIMO-VLC systems compared to traditional systems.

\section{Advanced Mathematical Tools for NOMA: From Optimization To ML}

An important design aspect that critically affects the performance of NOMA communication systems is the allocation of the available resources to the users. The resulting resource allocation problem can be handled either by utilizing conventional mathematical optimization methods or by leveraging new ML tools. In this section, we survey advanced resource allocation methods for NOMA from both the mathematical optimization and the ML perspectives.

\subsection{A Mathematical Optimization Perspective}
First, we survey the resource allocation methods that have been developed based on mathematical optimization theory. Generally speaking, resource allocation optimization problems can be categorized into two classes: Convex problems and non-convex problems \cite{convex,convex_analysis}, which can be dealt with by using convex and non-convex optimization methods, respectively, as detailed in the following. In Table \ref{table_optimization}, we summarize these methods as well as the related research contributions on NOMA.
\subsubsection{Convex optimization methods}
For convex resource allocation problems, where the objective is to minimize a convex function or to maximize a concave function, subject to a set of convex constraints \cite{convex}, there are generally two types of methods for finding the optimal solution: numerical methods and analytical methods. By making use of the convexity of the problem, both types of methods can yield the globally optimal solution with polynomial time complexity in general \cite{06_JSAC_Luo}.
\begin{itemize}	
	\item \emph{Numerical methods:} Various systematic numerical methods have been established for optimally solving convex optimization problems, including the Newton's method for unconstrained or equality-constrained convex optimization problems and the interior-point method for constrained convex optimization problems \cite{convex}. For example, the interior-point method was applied for solving the offloading ratio and uplink transmission time in a NOMA-MEC system \cite{18_CL_Chen}. Moreover, convex optimization problems can also be solved using off-the-shelf optimization toolboxes, e.g., CVX \cite{cvx} and YALMIP \cite{YALMIP}.		
	\item {\emph{Analytical methods:}} On the other hand, for convex optimization problems with a tractable structure, the optimal solution can be obtained analytically (e.g., in closed form or semi-closed form) by leveraging the \emph{Lagrangian duality method}, which can provide more insights than numerical methods. Specifically, under certain mild constraint qualification constraints (e.g., the Slater's condition), strong duality holds between a convex problem and the corresponding Lagrange dual problem \cite{convex}. Therefore, the optimal solution can be obtained by solving a set of equations under problem feasibility constraints, namely the Karush-Kuhn-Tucker (KKT) conditions; moreover, the optimal solution can be obtained by solving the dual problem, which may have a more tractable structure. Analytical methods have been applied for solving various convex resource allocation problems (especially power allocation problems) in NOMA networks. For example, by examining the KKT conditions, the authors of \cite{8352643} derived the optimal power allocation for a single-cell downlink NOMA network; the authors of \cite{19_WCL_Fang} obtained the optimal power allocation in an uplink multi-carrier NOMA system with user fairness considerations; the authors of \cite{20_TCOM_Fang} derived the optimal task partition ratios and offloading powers in closed form for a NOMA-MEC network; and the authors of \cite{20_TWC_Sena} derived the optimal dynamic power allocation for maximizing the worst-case user rate of a massive MIMO-NOMA network.		
	\item {\emph{Equivalent transformation from non-convex to convex problems:}} Besides problems that are originally in a convex form, convex optimization methods can also help to find the optimal solutions of non-convex problems that can be equivalently transformed into convex problems via manipulations, such as a change of variables, introduction of auxiliary variables, replacing the objective/constraint function with more tractable functions that are monotonic in terms of the original ones, etc. It is worth noting that the transformed problem must be \emph{equivalent} to the original one so that the optimal solution to the new convex problem can serve as the optimal solution to the original non-convex one. A typical example of such transformations is the class of geometric programming (GP) problems, where both the objective function and the constraints are posynomials (i.e., sum of monomials), thus the original problem is non-convex. However, by taking the logarithm of the objective function and constraints, this problem can be equivalently transformed into a convex optimization problem. Based on this favorable feature, various power control problems in multi-user networks can be approximated as GPs in the high-SNR regime and solved using convex optimization (a detailed tutorial is given in \cite{07_TWC_Chiang}). Using this approach, the power control problem has been handled for multi-carrier NOMA networks \cite{16_TWC_Di}, secure NOMA networks \cite{18_JSAC_Zhang}, and cognitive OFDM-NOMA networks \cite{18_TWC_Xu}.		
\end{itemize}
\subsubsection{Non-convex optimization methods}
Due to the diverse user demands and complicated network structures, most resource allocation problems are non-convex and cannot be equivalently transformed into convex optimization problems. In the following, we survey several useful non-convex optimization methods that are suitable for handling this class of resource allocation problems.
\begin{itemize}
	\item \emph{Matching theory}: Matching theory has recently emerged as a popular tool for solving the \emph{sub-channel allocation} problem in multi-carrier based NOMA networks. Specifically, unlike conventional OFDMA networks where each sub-channel is only assigned to at most one user to guarantee user channel orthogonality. NOMA enables power-domain multiplexing of multiple users on one sub-channel, which results in inter-user interference. Therefore, allocating multiple sub-channels to multiple users with the possibility of applying SIC in each sub-channel is much more challenging than the sub-channel allocation for OFDMA. Moreover, since the mapping between users and sub-channels is modeled using binary variables, such problems are combinatorial in nature, and cannot be solved using other conventional convex optimization methods either.
			
	Nevertheless, by identifying that the sub-channels and users can be viewed as two sets of selfish and rational players that interact with each other according to a given utility function (e.g., sum rate), the sub-channel allocation problem in NOMA can be modeled as a many-to-many \emph{matching game}, for which low-complexity algorithms can be designed by leveraging the matching theory \cite{matching}. Along this line, the authors of \cite{16_TCOM_Fang} proposed a greedy matching algorithm between sub-channels and users, where each sub-channel was assigned to two users. The authors of \cite{16_TWC_Di} considered the case where an arbitrary maximum number of users can be multiplexed on each sub-channel, and proposed a matching algorithm that converges to a stable matching. The authors of \cite{18_TWC_Liu} proposed a super-modular game based matching algorithm for maximizing the EE. Considering a system with mixed traffic types, the authors of \cite{20_TCOM_Youssef} devised a new matching-based sub-band allocation algorithm for both centralized and distributed antenna systems.
			
	Besides for sub-channel allocation, matching theory can also be applied for solving various other allocation/scheduling/assignment problems involving binary optimization variables \cite{16_SPM_Bayat,7878674}. Specifically, considering a MIMO-NOMA network operating in the mmWave band, the authors of \cite{18_TWC_Cui} proposed a matching algorithm for allocating different spatial mmWave beams to multiple users (also called \emph{user scheduling}), where each beam can be shared by an arbitrary number of users. The authors of \cite{20_TWC_Ding} developed a two-sided coalitional matching approach to jointly optimize the user clustering and BS selection. The authors of \cite{19_TVT_Zakeri} utilized matching theory to design codebook assignments for SCMA. Moreover, matching theory has also been shown effective for solving discrete resource allocation problems in various application scenarios, such as secure communications \cite{18_JSAC_Zhang}, cognitive networks \cite{18_TWC_Xu}, device-to-device (D2D) networks \cite{17_TCOM_Zhao,19_TCOM_Baidas}, and fog networks \cite{18_WC_Zhang,19_JSTSP_Wen}.
	\item \emph{Successive Convex Approximation (SCA)}: SCA is a low-complexity method to find a high-quality suboptimal solution to a non-convex problem with a non-convex objective function and/or non-convex constraints \cite{SCA}. Specifically, by finding convex tight upper bounds of the objective function and/or the constraint functions (which are referred to as the surrogate functions) at each local point, SCA iteratively solves the problem of minimizing this convex surrogate function with the constraint functions replaced by their surrogates, by using convex optimization methods, until convergence is reached. The surrogate function can be designed based on the form of the objective function or the constraint function (e.g., a common surrogate function for a concave objective/constraint function is its first-order Taylor series expansion). The SCA method has been widely used for optimizing the power allocation among multiple users/sub-channels, e.g., in outage probability constrained MIMO-NOMA networks \cite{18_TWC_Cui2}, FD multi-carrier NOMA networks \cite{17_TCOM_Sun}, mmWave-NOMA networks \cite{18_TWC_Cui}, energy-efficient NOMA networks \cite{18_TWC_Liu}, multi-user MIMO-NOMA networks \cite{18_TWC_Sun}, NOMA virtualized wireless networks \cite{18_WCL_Sinaie}, D2D-enabled NOMA networks \cite{19_Access_Chen}, and hybrid NOMA networks \cite{19_TVT_Wang}. Recently, the SCA method has also been shown effective for dealing with beamforming/trajectory design problems in UAV communications (see, e.g., \cite{19_TWC_Liu2,19_TCOM_Liu}) and RIS reflection coefficient optimization problems (see, e.g., \cite{20_WCL_Yang,20_TCOM_Yang}). Along this line, there are several related research contributions for NOMA systems, including \cite{19_ICC_Liu} for NOMA-based multi-beam UAV communications and \cite{Mu_IRS_NOMA,zhu_RIS_NOMA,21_TVT_Xie} for RIS-aided NOMA networks.	
	\item \emph{Branch-and-Bound (BnB):} Different from most other non-convex optimization methods that focus on finding a high-quality, yet generally suboptimal, solution with low computational complexity, the BnB method aims to obtain the globally optimal solution to a non-convex problem at the cost of high (often exponential) complexity, by systematically enumerating all the potentially optimal solutions \cite{66_OR_Lawler}. BnB is most commonly used for solving discrete optimization problems, and has later been extended to handling problems with continuous variables (see, e.g., \cite{BnB}). The basic idea of BnB is to partition the feasible set into smaller subsets, and find upper and lower bounds of the objective value for each subset. The partitioning and corresponding bounds are iteratively updated based on problem-specific rules, until convergence is reached. The BnB method has been applied for solving various resource allocation problems in wireless networks, such as the joint sub-channel, power, and rate assignment problem in OFDMA networks \cite{09_TWC_Lin} and the power control problem for maximizing the weighted sum rate in SISO interference channels \cite{11_TSP_Weeraddana}. Under the NOMA setup, the authors of \cite{17_TCOM_Wei} obtained the optimal resource allocation for a multi-carrier NOMA network with imperfect CSI by leveraging BnB; the authors of \cite{18_TWC_Cui} showed that the power allocation problem in a NOMA-mmWave network can be viewed as a BnB problem, and obtained the $\epsilon$-optimal power allocation policy.	
	\item \emph{Monotonic Optimization:} Besides BnB, monotonic optimization can also find the globally optimal solution of a non-convex optimization problem. Specifically, it is suitable for problems whose objective and constraints are monotonic (i.e., strictly increasing or decreasing) in the optimization variables. By making use of this monotonicity, the search of the optimal solution within the feasible set can be made more efficient (more details can be found in \cite{12_Zhang}). The effectiveness of monotonic optimization has made it appealing for obtaining the globally optimal solution to several well-known difficult problems such as the power control problem in ad hoc networks \cite{09_TWC_Qian} and the weighted sum rate maximization problem for general $K$-user MISO/single-input multiple-output (SIMO) Gaussian interference channel \cite{12_TWC_Liu}. Employing monotonic optimization, the authors of \cite{17_TCOM_Sun} obtained the optimal joint sub-carrier and power allocation strategy for an FD multi-carrier NOMA system and the authors of \cite{8307182} designed the optimal power control in NOMA relay-assisted networks.
	\item \emph{Block Coordinate Descent (BCD)}: The methods introduced above are mostly suitable for optimizing one particular type of resources, e.g., power control or sub-channel allocation. However, to optimize the overall network performance, different types of resources need to be jointly optimized, which results in more complicated problems with potentially various types of variables coupled in the objective function and/or constraints. Instead of directly solving such joint optimization problems, which is typically difficult, BCD (sometimes also referred to as \emph{alternating optimization}) is a favorable alternative which groups the optimization variables into multiple blocks (e.g., grouping the power control variables into one block, and the sub-channel allocation variables into another block), and iteratively optimizes each block of variables with the other blocks being fixed. In this way, the joint optimization problem is decoupled into several sub-problems each over a smaller group of variables, which can be handled using the methods described above. It is worth noting that despite the typically low complexity of the BCD method, the algorithm convergence and the solution quality need to be judiciously evaluated, by checking the optimality of the solution obtained for each subproblem and examining the specific problem structure. The BCD method has been widely applied for solving resource allocation problems in NOMA networks. For example, the authors of \cite{19_JSTSP_Zeng} optimized the uplink and downlink power allocation iteratively using BCD for SE maximization; for a beamspace MIMO-NOMA system, a BCD algorithm was proposed for iteratively optimizing the power allocation and beam-specific digital precoding \cite{20_TSP_Jiao}; to characterize the capacity region of an RIS-aided network, the authors of \cite{shuowen_capacity} considered NOMA transmission and developed a BCD algorithm for iteratively optimizing each RIS reflection coefficient, where the optimal solution to each sub-problem was obtained.
	\item \emph{Problem-specific methods via structure exploitation}: Although each of the above methods can be applied for solving a certain type of problem, there are still many problems that cannot be easily dealt with using these methods, due to the intractability of the objective function and/or constraints (e.g., the unit-modulus constraints in hybrid/analog/constant-envelope precoding problems \cite{16_JSTSP_Yu,16_WCL_Zhang,18_TCOM_Zhang} and reflection coefficient design in RIS-aided networks). In this case, by carefully exploiting the specific structure of a problem, hidden information on the optimal solution may be unveiled, which facilitates the design of problem-specific methods that can find a high-quality solution with low complexity. Moreover, note that even for the case where the above methods can be applied, problem-specific methods may yield improved performance with lower complexity since they make full use of the problem structure. For example, for the non-convex OFDMA sub-channel allocation problem, the authors of \cite{06_TCOM_Yu} unveiled that under a time-sharing condition which holds for a large total number of sub-channels, the duality gap between the primal and dual problems is zero. Hence, this non-convex problem can be equivalently solved by solving the dual problem, where the variables are not coupled and the optimal solution can be easily found. By exploiting the unique channel structure of an RIS-aided MIMO system, the authors of \cite{20_JSAC_Zhang2} derived the optimal solution for the RIS reflection coefficients in closed form, despite the non-convexity of the unit-modulus constraint. For a two-user RIS-aided NOMA network, the authors of \cite{shuowen_capacity} revealed that the optimal solution for the RIS reflection coefficients can be obtained by solving a relaxed convex optimization problem, where the relaxation is tight. For the more complicated resource allocation problems arising in the context of NGMA, we expect problem-specific methods to play a more important role due to their capability of extracting insights for complex problems.
\end{itemize}
\begin{table*}[t]\large
\caption{Summary of mathematical optimization methods applied in NOMA.}
\begin{center}
\centering
\resizebox{\textwidth}{!}{
\begin{tabular}{!{\vrule width1.5pt}c!{\vrule width1.5pt}c!{\vrule width1.5pt}c!{\vrule width1.5pt}c!{\vrule width1.5pt}c!{\vrule width1.5pt}}
\Xhline{1.5pt}
\centering
\makecell[c]{\textbf{Optimization Category}}  & \makecell[c]{\textbf{Optimization Method}} &\makecell[c]{\textbf{Characteristics}} & \makecell[c]{\textbf{Optimization Variable Type}} & \makecell[c]{\textbf{Ref.}}\\
\Xhline{1.5pt}
\centering
\multirow{5}{*}{\makecell[c]{Convex optimization method}} & {\makecell[c]{Numerical methods}} & {\makecell[c]{Globally optimal solution\\ Polynomial complexity}} & Continuous & \cite{18_CL_Chen} \\
\cline{2-5}
\centering
&  {\makecell[c]{Analytical methods}}  & {\makecell[c]{Globally optimal solution\\ Polynomial complexity}} & Continuous & \cite{20_TWC_Sena}, \cite{8352643}, \cite{19_WCL_Fang}, \cite{20_TCOM_Fang}, \\
\cline{2-5}
\centering
& {\makecell[c]{Equivalent transformation \\ from non-convex to convex problems}} & {\makecell[c]{Globally optimal solution\\ Polynomial complexity}} & {\makecell[c]{Continuous \\(the transformed convex problems)}} & \cite{16_TWC_Di}, \cite{18_JSAC_Zhang}, \cite{18_TWC_Xu}\\
\Xhline{1.5pt}
\centering
\multirow{11}{*}{\makecell[c]{Non-convex optimization method}}&   Matching theory & {\makecell[c]{Generally suboptimal solution\\ Low complexity}}  & Discrete & \cite{16_TCOM_Fang,16_TWC_Di,18_TWC_Liu,20_TCOM_Youssef,18_TWC_Cui,20_TWC_Ding,19_TVT_Zakeri,18_JSAC_Zhang,18_TWC_Xu,17_TCOM_Zhao,19_TCOM_Baidas,18_WC_Zhang,19_JSTSP_Wen} \\
\cline{2-5}
\centering
&  SCA  & {\makecell[c]{Generally suboptimal solution\\ Low complexity}} & Continuous & {\makecell[c]{\cite{17_TCOM_Sun,18_TWC_Cui,18_TWC_Liu,18_TWC_Cui2,18_TWC_Sun}\\\cite{18_WCL_Sinaie,19_Access_Chen,19_TVT_Wang,19_ICC_Liu,Mu_IRS_NOMA,zhu_RIS_NOMA,21_TVT_Xie}}}\\
				\cline{2-5}
				&  BnB  & {\makecell[c]{Globally optimal solution\\ High (often exponential) complexity}} & Discrete/continuous & \cite{17_TCOM_Wei,18_TWC_Cui}\\
\cline{2-5}
\centering
& Monotonic optimization & {\makecell[c]{Globally optimal solution\\ High (often exponential) complexity}} & Discrete/continuous & \cite{17_TCOM_Sun}\\
\cline{2-5}
\centering
& BCD & {\makecell[c]{Generally suboptimal solution\\ Low complexity}} & Continuous/discrete & \cite{19_JSTSP_Zeng,20_TSP_Jiao,shuowen_capacity}\\
\cline{2-5}
\centering
& {\makecell[c]{Problem-specific methods \\via structure exploitation}} & {\makecell[c]{Generally high-quality or even optimal solution\\Low complexity}} & Continuous/discrete & \cite{shuowen_capacity}\\
\Xhline{1.5pt}
\end{tabular}
}
\end{center}
\label{table_optimization}
\end{table*}

\subsection{An ML Perspective}
ML techniques have gained increasing popularity for resource allocation in heterogeneous wireless networks due to their learning capability, especially in time-varying and unpredictable network environments~\cite{liu2019machine,sun2019application,chen2019artificial}. In this subsection, we will provide an overview of the ML tools employed in NOMA communication, see also Table \ref{table:ML}.
\subsubsection{Deep learning for NOMA}
Due to its capability of accurately tracking the state of networks and then predicting their future evolution, deep learning (DL) has been applied in NOMA to solve diverse problems such as acquiring perfect CSI~\cite{gui2018deep}, signal constellation design~\cite{jiang2019deep}, signal detection~\cite{xie2020deep,emir2021deep}, user clustering~\cite{dejonghe2021deep}, and resource allocation~\cite{sun2019deep,huang2020deep,saetan2019power,yang2019deep,ali2021deep,fu2019dynamic,lin2019deep}. The core idea of employing DL approaches in NOMA is that the DL model uses a deep neural network (DNN) to find a suitable data representation in each layer. On the one hand, when solving problems such as acquiring timely and accurate CSI, the DL approach in NOMA can be data-driven, where the NOMA communication system is optimized for a large training dataset. On the other hand, when solving problems such as resource allocation, the DL approach in NOMA can be model-based, where some prior knowledge of the NOMA communication system is exploited to reduce the number of parameters to be learned. Since the performance of NOMA highly depends on the adopted resource allocation strategy, in the following, we mainly focus on existing research contributions on DL-empowered resource allocation for NOMA communication systems.\\
\indent In an effort to find the optimal decoding order and power allocation policy in an NOMA-enhanced satellite-based IoT (S-IoT) scenario, the authors of~\cite{sun2019deep} proposed a DL-based long-term power allocation (DL-PA) algorithm, which determines the desired decoding order based on the queue state and the channel state of the users. The obtained results revealed that the proposed DL-based algorithm outperforms other benchmarks in terms of long-term network utility, average arriving rate, and queuing delay. The authors of~\cite{huang2020deep} proposed a DNN-based algorithm to maximize the sum data rate and EE in MIMO-NOMA networks. By leveraging the proposed algorithm, the challenges caused by the rapidly changing channel conditions and the extremely complex spatial structure limitations were tackled. To maximize the sum rate for a downlink NOMA system in the presence of imperfect SIC, the authors of~\cite{saetan2019power} proposed a DNN-based algorithm to predict the optimal power allocation through exhaustive search. Furthermore, a similar problem was investigated by the authors of~\cite{yang2019deep}, where a DNN-based method with a simple operation was proposed and can accomplish near real-time resource allocation. To maximize the sum rate in NOMA enabled D2D systems, the authors of~\cite{ali2021deep} proposed a DNN framework to determine a joint power loading solution at both the source and relaying nodes. It was shown that the proposed DNN-based algorithm outperforms the conventional convex optimization-based approach in terms of sum rate and computational complexity. As a further step, the authors of~\cite{fu2019dynamic} proposed a dynamic power control scheme for NOMA enhanced wireless caching networks, where a DNN-based algorithm was proposed to strike a balance between performance and the computational complexity. The authors of~\cite{lin2019deep} studied NOMA-based wireless power transfer networks, where the harvested energy and the transmission rate were optimized subject to the requirements of energy harvesting and communication, respectively. A deep belief network (DBN)-based algorithm, which comprises preparing data samples, training, and running, was proposed to solve the formulated problem.
\subsubsection{Reinforcement learning for NOMA}
Since the conventional mathematical optimization approaches are heavily system-dependent, they are computationally expensive and incur considerable timing overhead, and they may fail to be accurate when facing rapidly time-varying wireless systems or environments~\cite{Xiao2021Artificial}. Reinforcement learning (RL), which empowers the models to learn from the dynamic/uncertain environment and to learn from their historical experiences, is capable of overcoming these challenges and provides an efficient approach for tackling resource allocation problems in dynamic heterogeneous networks~\cite{luong2019applications}. The core idea of employing RL approaches in NOMA is that RL enables the AP to interact with the environment. In each time slot, the AP periodically observes the state of the NOMA communication system. The state space consists of the wireless resources (e.g., power resource, bandwidth resource, computing resource, and caching resource) allocated to each user. An action is carried out by the AP to select the optimal control policy, which allows the RL model to obtain the maximum Q-value in each time slot. By carrying out an action, the AP will receive a penalty/reward, which is determined by the considered objective function. In general, the principle of RL algorithms is to maximize the long-term reward. RL algorithms can be divided into three categories, namely, value-based algorithms, policy-based algorithms, and actor-critic algorithms. In addition, algorithms can be categorized based on the state of the model, i.e., discrete or continuous.\\
\indent RL approaches have been widely employed for solving resource allocation problems in NOMA networks~\cite{he2019joint}. The authors of~\cite{xiao2017reinforcement} proposed a hotbooting Q-learning based NOMA power allocation algorithm for downlink NOMA transmission, where the jamming and radio channel parameters do not have to be known. The proposed algorithm was shown to outperform the standard Q-learning-based strategy. By formulating a long-term throughput maximization problem as a Partially Observable Markov Decision Process (POMDP), the authors of~\cite{zhang2020deep} developed a Long Short-Term Memory (LSTM) based Deep Q-network (LSTM-DQN) algorithm to solve the formulated problem and reduce collisions in the frequency domain as well as the computational complexity of the DQN algorithm. To carry out the real-time channel assignment in a real-time dynamic environment, the authors of~\cite{zheng2021channel} proposed an actor-critic based algorithm for a hybrid NOMA network. The results showed that the proposed algorithm has superior environmental adaptability with low time complexity. Furthermore, based on the Deep Deterministic Policy Gradient (DDPG) algorithm, the authors of~\cite{wang2021joint} proposed a deep reinforcement learning (DRL)-based joint resource management algorithm to maximize the weighted-sum system throughput of a practical multi-carrier NOMA (MC-NOMA) system.\\
\indent Besides the above conventional NOMA communication systems, RL-empowered NOMA has been also extended to diverse application scenarios towards 6G. For instance, in the context of UAV aided communications, the authors of \cite{khairy2020constrained} applied NOMA to enhance the massive channel access in a UAV-aided IoT network. Based on Lagrangian primal-dual policy optimization, an online model-free constrained DRL algorithm was proposed to maximize the total network capacity under stochastic mutual interference from IoT devices. The authors of~\cite{zhong2020multi} jointly optimized the three-dimensional trajectory of multiple UAVs and the power allocation policy to maximize the total throughput in a UAV-BS enabled NOMA network. A Mutual Deep Q-network (MDQN) algorithm was proposed to solve the formulated problem. The algorithm was shown to have a faster convergence speed than the conventional DQN algorithm in the multi-agent case. In the context of NOMA-MEC networks, the authors of~\cite{yang2020cache} proposed a Single-Agent Q-learning (SAQ-learning) algorithm and a Bayesian Learning Automata (BLA) based Multi-Agent Q-learning (MAQ-learning) algorithm to obtain a long-term resource allocation strategy and task offloading decisions, respectively. They showed that the proposed NOMA-MEC scheme significantly outperforms other benchmark schemes, such as all-local computing and all-offloading computing. The authors of~\cite{li2020resource} proposed a Mean-Field DDPG (MF-DDPG)-based algorithm to solve the resource allocation problem for NOMA-MEC in an ultra-dense network. Numerical results showed that the proposed algorithm can efficiently reduce both the energy consumption and task delay of users. For an RL-empowered RIS-NOMA communication system, the authors of~\cite{Xiao2021RIS} proposed a decaying double deep Q-network (${{\text{D}}^{\text{3}}}{\text{QN}}$) based position-acquisition and phase-control algorithm to attain the optimal deployment and design policies for the RIS as well as the NOMA power allocation policy. The authors of~\cite{yang2021machine} proposed a DDPG-based algorithm to jointly optimize the NOMA user partition, RIS phase shift, and NOMA power allocation policy to maximize the sum rate. In contrast to the conventional training-then-testing procedure, the proposed algorithm can obtain a dynamic resource allocation policy based on a long-term self-adjusting learning model.
\subsubsection{Federated learning for NOMA}
Federated learning (FL), which can achieve distributed network training by only exchanging network parameters (e.g., weights, gradients) rather than exchanging the sensitive and private raw data~\cite{niknam2020federated,chen2020joint}, is also regarded as a promising technique to tackle problems in distributed application scenarios of NOMA. In the FL model, since the local data is not necessary for global FL training and only the parameters of the local networks are needed by the central network, the privacy of the users can be well protected.\\
\indent Motivated by the above unique advantages, some initial research has begun to study the benefits of employing FL in NOMA communication systems. The authors of~\cite{habachi2019fast} studied a NOMA-enabled machine type devices (MTDs) communication system, where an FL approach was proposed to allow the BS and MTDs to collaboratively estimate the traffic model, thus improving the NOMA capacity. The FL concept was further studied by the authors of~\cite{zhong2021mobile}, where an indoor RIS-aided NOMA communication system was considered. In particular, an FL-enabled DRL algorithm was proposed to accelerate the training process for the joint optimization of the deployment location and phase shifts of the RIS and the power allocation of the NOMA users. The result verified that the proposed FL-enabled algorithm can achieve a faster convergence rate and a better optimization performance compared to an independent training scheme.

\subsubsection{Other machine learning techniques for NOMA}
Apart from the aforementioned ML approaches, some other machine learning algorithms, such as transfer learning and meta-learning, are also promising for adoption to further improve the performance of the existing ML algorithms. Additionally, some supervised/unsupervised learning algorithms can also be invoked to solve problems in NOMA communication systems with a low complexity.

\begin{itemize}
    \item \emph{Transfer learning for NOMA}: The authors of~\cite{zhong2020path} proposed a deep transfer deterministic policy gradient (DT-DPG) algorithm to maximize the mission efficiency and communication reliability of an indoor RIS-aided NOMA communication system. In particular, the radio map was invoked as a virtual environment to train the RL agent, which can significantly reduce the training time and hardware costs. In an effort to mitigate interference in NOMA-enhanced mm-Wave communications, the authors of~\cite{Elsayed2021Transfer} considered the joint user-cell association and beam selection problem to maximize the total network capacity. A transfer Q-learning algorithm was proposed to solve the formulated problem. It was shown that the proposed transfer Q-learning algorithm outperforms the benchmarks in dynamic scenarios and also converges faster than the conventional Q-learning algorithm.
    \item \emph{Supervised/Unsupervised learning for NOMA}: Supervised learning algorithms, such as the support vector machine (SVM) algorithm~\cite{chen2015support}, have been adopted for user mobility/position prediction in NOMA networks with mobile nodes. Unsupervised learning algorithms, such as the principal component analysis (PCA) algorithm~\cite{gkonis2020non} and the K-means algorithm~\cite{cui2018unsupervised,gao2021machine,zhang2020energy}, have been used for user clustering in NOMA networks.
\end{itemize}

\begin{table*}[htbp]\large
\caption{Summary of existing research contributions on ML-empowered NOMA}
\begin{center}
\centering
\resizebox{\textwidth}{!}{
\begin{tabular}{!{\vrule width1.5pt}c!{\vrule width1.5pt}c!{\vrule width1.5pt}c!{\vrule width1.5pt}l!{\vrule width1.5pt}l!{\vrule width1.5pt}}
\Xhline{1.5pt}
\centering
 \makecell[c]{\textbf{Ref.}} &\makecell[c]{\textbf{Scenarios}} & \makecell[c]{\textbf{Algorithm}}  & \makecell[c]{\textbf{Main Objectives}} & \makecell[c]{\textbf{Characteristics}} \\
\Xhline{1.5pt}
\centering
 \cite{sun2019deep} & NOMA S-IoT & DL-PA  &  Network utility &  Long-term power allocation\\
\hline
\centering
 \cite{huang2020deep} & MIMO-NOMA & CDNN  & EE &  Several convolutional layers and multiple hidden layers \\
\hline
\centering
 \cite{saetan2019power,yang2019deep} & Multi-user NOMA & DNN   & Sum rate &  Imperfect successive interference cancellation\\
\hline
\centering
\cite{ali2021deep} & NOMA-D2D & DNN   & Sum rate & Channel coefficients, power budget, and binary user pairing variable serve as DNN input\\
\hline
\centering
\cite{fu2019dynamic} &  Multi-user NOMA & DNN   & Transmission delay &  Dynamic power control\\
\hline
\centering
\cite{lin2019deep} &  NOMA-SWIPT & DBN  & Rate and harvested energy  & Comprises three phases: Preparing data samples, training, and running\\
\hline
\centering
\cite{xiao2017reinforcement} &  Multi-user NOMA & Hotbooting Q-learning  & Sum rate & Without relying on the knowledge of the jamming and radio channel parameters\\
\hline
\centering
\cite{zhang2020deep} &  Grant-free NOMA & LSTM-based DQN  & Throughput & The long-term cluster throughput maximization
problem is formulated as a POMDP problem\\
\hline
\centering
\cite{zheng2021channel} &  Hybrid NOMA & Actor-critic  & Sum rate & The proposed approach includes recurrent neural network (RNN) and proximal policy optimization (PPO)\\
\hline
\centering
\cite{wang2021joint} &  MC-NOMA & DDPG-based DRL-JRM  & Weighted sum rate & A novel centralized action-value function is designed to measure the reward\\
\hline
\centering
\cite{khairy2020constrained} &  NOMA-UAV & CDRL  & Network capacity & The proposed algorithm is based on PPO algorithm\\
\hline
\centering
\cite{zhong2020multi} &  NOMA-UAV & MDQN  & Throughput & The proposed algorithm has a faster convergence rate than the conventional DQN algorithm in the multi-agent case\\
\hline
\centering
\cite{yang2020cache} &  NOMA-MEC & BLA-based MAQ-learning  & Energy consumption & The BLA based action selection scheme is adopted for every agent to obtain optimal action in every state\\
\hline
\centering
\cite{li2020resource} &  NOMA-MEC & MF-DDPG & Energy consumption, delay & The performance of the proposed algorithm is better than that of DQN\\
\hline
\centering
\cite{Xiao2021RIS} &  NOMA-RIS & ${{\text{D}}^{\text{3}}}{\text{QN}}$  & EE& Users' dynamic tele-traffic demand is considered by leveraging a real dataset\\
\hline
\centering
\cite{yang2021machine} &  NOMA-RIS & DDPG  &  Sum rate & The proposed algorithm can learn the dynamic resource allocation policy using a long-term self-adjusting learning model\\
\hline
\centering
\cite{habachi2019fast} &  NOMA-MTD & FL & QoS & Allocating different transmit powers for an MTD in
alarm mode and regular mode\\
\hline
\centering
\cite{zhong2021mobile} &  NOMA-RIS & FL  & Sum rate & FL is adopted to enable multiple agents to simultaneously
explore similar environments and exchange experiences\\
\hline
\centering
\cite{zhong2020path} &  NOMA-RIS & DT-DPG  &  Mission quality indicator & Map is invoked as a virtual environment to train the RL agent\\
\hline
\centering
\cite{Elsayed2021Transfer} &  NOMA-mmWave & Transfer Q-learning &  Sum rate & The proposed algorithm converges faster than Q-learning algorithm\\
\Xhline{1.5pt}
\end{tabular}
}
\end{center}
\label{table:ML}
\end{table*}

\subsection{Discussion and Outlook}
Based on the above discussion, it can be observed that both mathematical optimization and ML approaches can be effective for solving diverse problems arising in the context of NOMA in different scenarios. In this subsection, we discuss both advantages and disadvantages of mathematical optimization and ML approaches, and identify principles for selecting suitable algorithms.
\begin{itemize}
    \item \emph{Mathematical optimization or ML approaches?} Next-generation wireless networks are more complicated due to their large-scale, versatile, and heterogeneous nature~\cite{RIS_survey,alwarafy2021deep}. Conventional mathematical optimization methods require complete or quasi-complete knowledge of the wireless environment, which is non-trivial to obtain in heterogeneous scenarios. In addition, conventional mathematical optimization generally aims for instantaneously achieving minimum benefits for networks without having an architectural vision in long-term and dynamic scenarios. Hence, ML approaches outperform conventional mathematical optimization when encountering complex time-variant hybrid environments or even unknown environments. Moreover, in this case, the formulated NOMA optimization problems usually have to compromise between multiple objectives (e.g., delay, throughput, transmit power) in order to obtain a good solution. This further makes conventional mathematical optimization schemes inefficient, since determining the entire Pareto-front of optimal solutions is challenging for them. In particular, the search-space is expanded as the number of parameters increases, which makes conventional gradient-based optimization techniques unsuitable~\cite{wang2020thirty}. However, when faced with static or quasi-static scenarios, ML algorithms generally are not superior since their optimality cannot be theoretically proved or strictly guaranteed, while that of the conventional mathematical optimization can. In addition, ML approaches also loose their advantages in terms of complexity in static environments due to the required training.

    \item \emph{DL or DRL?} Although both DL and DRL are widely employed approaches in NOMA applications~\cite{gao2021machine,DLchannel,jiang2020channel,DLbeamforming}, their performance, applicability, and respective advantages have not been investigated thoroughly, yet. On the one hand, a notable feature of DRL is its benefits in long-term optimization~\cite{Xiao2021Artificial,Xiao2021RIS,zhong2020path,Xiao2019Trajectory}. However, some research contributions have demonstrated that DL algorithms can also enable long-term optimization by focusing not only on short-term benefits in resource allocation~\cite{sun2019deep}. DRL is generally used for problems which can be formulated as Markov decision processes (MDPs), which means that, for a system whose state can be changed and adjusted arbitrarily without being restricted by the state in the previous time slot, such as transmit power optimization, channel estimation, etc., DRL approaches loose their advantages. On the contrary, for MDP problems, such as UAV trajectory design subject to flight speed and long-term resource allocation with varying power constraints, DRL has preeminent potential compared to DL. On the other hand, in contrast to DL, DRL can be trained by interacting with the environment, which does not require any labeled dataset. There are two key steps: {\em training} and {\em inference}. During the training phase, the agent interacts with the environment to collect experience. These collected experiences constitute the training dataset of the RL agent and will be used to learn the optimal decision-making rule. The environment is a simulator mimicking the real-world system. In data-driven systems, DL may be advantageous over DRL, while DRL performs better in model-driven systems. It is worth noting that DRL can be used even when the ``model'' is unknown (e.g., ``model-free'' RL/DRL methods). Both single and multi-agent DRL methods are being explored for designing future AI-enabled wireless networks~\cite{9372298}.
\end{itemize}

\section{Road Ahead: A Multi-Antenna and NOMA-Based Unified Framework for NGMA}

From the discussion in the previous sections, we can observe that NOMA is expected to play an important role in NGMA. On the one hand, there are promising application scenarios in 6G, where NOMA can be employed for further performance enhancement (c.f., Section IV), and emerging physical layer techniques, which NOMA can interplay with to create win-win transmission strategies (c.f., Section V). On the other hand, there are advanced mathematical tools for enabling NOMA to overcome the stringent requirements and challenges in next generation wireless networks (c.f., Section VI). Motivated by these considerations, in this section, we present a unified framework for NGMA based on multi-antenna transmission and NOMA, including both downlink and uplink multi-user communication scenarios. For each scenario, we discuss the corresponding signal models and implementation principles. The aim of the proposed unified NGMA framework is to fully reap the benefits of multiple antenna and NOMA techniques and to enable highly flexible and customized communication services for future wireless networks. To begin with, we first discuss issues arising from the use of SIC in MIMO-NOMA, which may cause NOMA to perform worse than SDMA when the multi-antenna communication system is underloaded/critically loaded. This also provides one of the main motivations for developing advanced NGMA techniques.

\subsection{SIC Issues in MIMO-NOMA}
In NOMA, SIC is employed at the receivers to remove the inter-user interference caused by resource sharing. To ensure that SIC can be successfully carried out at the strong user, the decoding rate condition ${R_{w \to s}} \ge {R_{w \to w}}$ has to be satisfied~\cite{18_WC_Liu}, where ${R_{w \to s}}$ and ${R_{w \to w}}$ denote the achievable communication rates at the strong user and weak user when decoding the weak user's signal, respectively. This means that we need to keep the weak user's signal strength (i.e., the inter-user interference) received at the strong user at a moderate or a relatively high level to be perfectly decoded. It is also worth mentioning that, in some literature~\cite{16_TSP_Hanif,quasi-degradation}, the above decoding rate condition is replaced with a more relaxed condition, i.e., ${R_w} = \min \left( {{R_{w \to s}},{R_{w \to w}}} \right)$, which means that the achievable communication rate of the weak user is limited to the decoding rates achieved at the strong and weak users, respectively. Nevertheless, this relaxed condition requires the same principle as the original one, i.e., the strong user has to receive a considerable amount of power of the weak user's signal to carry out SIC.\\
\indent For single-antenna multi-user communication systems, this decoding rate condition can be automatically satisfied by ordering the users in terms of their scalar channel power gains, which, however, is not possible anymore for multi-antenna communication systems where the users' channels are vectors or matrices. More importantly, the spatial DoFs provided by multi-antenna techniques can be exploited to enhance the intended signal strength while mitigating the inter-user interference for each user, i.e., SDMA. Therefore, at first glance, one may think that SDMA is superior to NOMA in multi-antenna communications, since it enables a complete or almost interference-free transmission for all users over the same radio resource while NOMA only allows one user (i.e., the strongest user if the beamformer-based NOMA is employed) to receive the intended messages without inter-user interference. Moreover, employing NOMA in multi-antenna communication systems imposes additional SIC decoding constraints while SDMA does not. \emph{Does carrying out SIC make NOMA ineffective in multi-antenna communication systems?} To answer this question, we compare NOMA with SDMA for a downlink MISO communication system, where an $N$-antenna BS serves $K$ single-antenna users. Let ${{\mathbf{h}}_k^H}\in {{\mathbb{C}}^{1 \times {N}}}$ and ${{\mathbf{w}}_k}\in {{\mathbb{C}}^{N \times {1}}}$ denote the channel vector and the transmit beamformer of user $k \in \left\{ {1,2, \ldots ,K} \right\}$, respectively. In the following, we consider both the \emph{underloaded/critically loaded} (i.e., $N \ge K$) and \emph{overloaded} (i.e., $N < K$) regimes.

\subsubsection{Underloaded/Critically loaded system} In this case, for ease of presentation, we assume that $N \ge K=2$. In terms of the correlation between the two user channels, we consider the following two scenarios.\\
\indent \emph{{Scenario 1}}: Assume that ${{\mathbf{h}}_1^H}$ and ${{\mathbf{h}}_2^H}$ are mutually orthogonal. In this scenario, it is trivial to design two beamformers, ${{\mathbf{w}}_1}$ and ${{\mathbf{w}}_2}$, for the two users such that
\begin{align}\label{zf}
{\mathbf{h}}_1^H{{\mathbf{w}}_2} = {\mathbf{h}}_2^H{{\mathbf{w}}_1} = 0,
\end{align}
i.e., there is no inter-user interference at each user. However, for NOMA in this scenario, the user ordering and the SIC condition become unnecessary since neither of the two users has to decode the other's signal to remove the inter-user interference. Therefore, SDMA outperforms NOMA in this scenario.\\
\indent \emph{{Scenario 2}}: Assume that ${{\mathbf{h}}_1^H} = c{{\mathbf{h}}_2^H}$, where $c$ is a constant. In this scenario, as the two channel vectors are highly correlated\footnote{As discussed before, when the communication link is LoS dominated (e.g., UAV-aided communications and mmWave/THz communications), there is a high probability that the users' channels are highly correlated.}, it is impossible to design two beamformers satisfying \eqref{zf} even if the underloaded case is considered. As a result, with SDMA, both users will suffer from severe inter-user interference. Fortunately, with the assistance of SIC, NOMA guarantees that one user can remove the interference caused by the other user, instead of letting all users suffer from interference. Therefore, NOMA will outperform SDMA in this case. For scenario 2, one may argue that NOMA imposes additional SIC decoding constraints, which reduce the overall DoFs to be exploited. However, it is worth mentioning that for highly-correlated channel vectors, the resulting SIC decoding conditions are easy to satisfy.
\begin{figure*}[t!]
\begin{center}
    \includegraphics[width=4.5in]{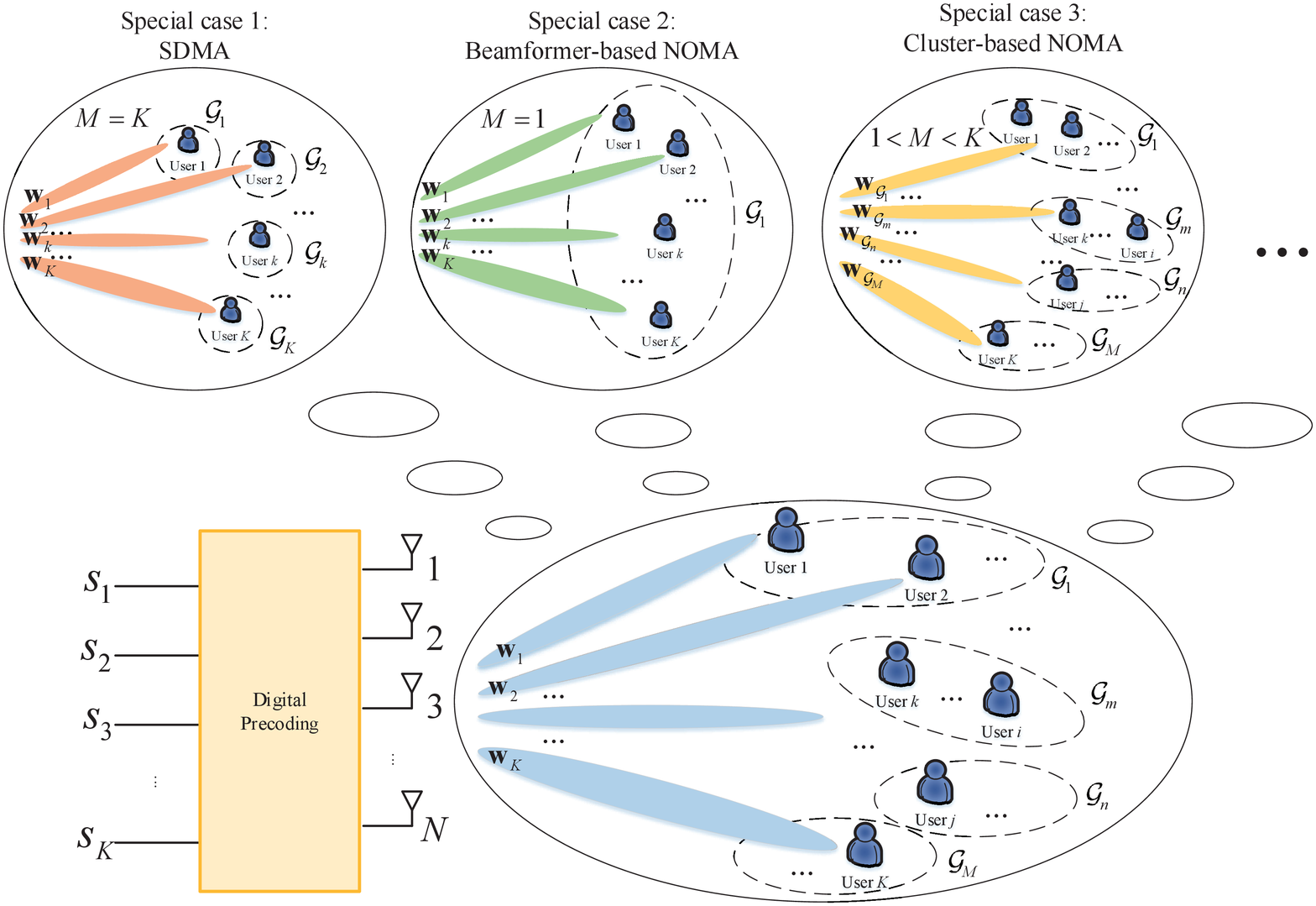}
    \caption{Illustration of the proposed multi-antenna and NOMA-based unified NGMA framework for downlink transmission.}
    \label{Downlink}
\end{center}
\end{figure*}
\subsubsection{Overloaded system} In this case, for ease of presentation, we assume that $N=2$ and $K=4$. Due to the insufficient spatial DoFs in the overloaded case, with SDMA, all users will suffer from severe inter-user interference if they are simultaneously served. In contrast, NOMA, especially cluster-based NOMA, is a promising transmission option in the overloaded case. To demonstrate this, we assume that the channels of users 1 and 3 are highly correlated (i.e., ${{\mathbf{h}}_1^H} = c_1{{\mathbf{h}}_3^H}$) and are grouped into cluster 1, while the channels of users 2 and 4 are highly correlated (i.e., ${{\mathbf{h}}_2^H} = c_2{{\mathbf{h}}_4^H}$) and are grouped into cluster 2. In this case, we can design two beamformers, ${{\mathbf{w}}_{c1}}$ and ${{\mathbf{w}}_{c2}}$, such that for the two clusters we have
\begin{subequations}\label{zf2}
\begin{align}
{\mathbf{h}}_2^H{{\mathbf{w}}_{c1}} &= {\mathbf{h}}_4^H{{\mathbf{w}}_{c1}} = 0,\\
{\mathbf{h}}_1^H{{\mathbf{w}}_{c2}} &= {\mathbf{h}}_3^H{{\mathbf{w}}_{c2}} = 0.
\end{align}
\end{subequations}
For each cluster, the superimposed signal of the two assigned users is transmitted via the same beamformer to carry out NOMA. By doing so, the strong NOMA user in each cluster can receive its intended signal in an interference-free manner since the inter-cluster interference is mitigated by exploiting the spatial DoFs and the intra-cluster interference is canceled by employing SIC. At the same time, two additional users can be simultaneously served, which would not be possible with SDMA. Therefore, in the overloaded regime, we can observe that the limited spatial DoFs can be exploited more efficiently with NOMA than with SDMA.\\
\indent Based on the above examples, we can conclude that for multi-antenna communication systems, on the one hand, NOMA may sometimes be ineffective compared to SDMA in the underloaded/cirtically loaded regime due to the improper employment of SIC. On the other hand, in the underloaded/critically loaded regime with highly correlated channels and in the overloaded regime, NOMA is indeed effective and SIC is essential to fully exploit the available spatial DoFs. Therefore, we cannot simply rely on one specific transmission scheme in multi-antenna communication systems. Moreover, given the fact that practical wireless environments are much more complicated than the above presented scenarios, it will not be possible to simply switch between different transmission strategies to cope with the diverse communication scenarios occurring in actual wireless systems. To address this issue, in the following, we propose a multi-antenna and NOMA based unified framework for NGMA, which aims to exploit the benefits of both techniques to facilitate more flexible transmission than current MIMO-NOMA and SDMA.

\subsection{A Unified Framework for NGMA: Downlink Case}
To begin with, as shown in Fig. \ref{Downlink}, we consider the downlink communication of a typical MISO multi-user communication system, where an $N$-antenna BS serves $K>1$ single-antenna users. Let ${\mathbf{s}} = {\left[ {{s_1},{s_2}, \ldots ,{s_K}} \right]^T}$ denote the information vector, where ${{s_k}}$ denotes the information symbol intended for user $k \in {\mathcal{K}} \triangleq \left\{ {1,2, \ldots ,K} \right\}$. Moreover, let ${{\mathbf{w}}_k} = \sqrt {{p_k}} {\overline {\mathbf{w}} _k}\in {{\mathbb{C}}^{{N} \times 1}}$ denote the corresponding beamformer for transmitting ${{s_k}}$ to user $k$, where ${\overline {\mathbf{w}} _k}$ and $p_k$ denote the normalized vector and the transmit power, respectively. Therefore, the transmit signal at the BS is given by ${\mathbf{x}} = \sum\limits_{k = 1}^K {{{\mathbf{w}}_k}{s_k}}$. The downlink channel vector between the BS and user $k$ is denoted by ${\mathbf{h}}_k^H \in {{\mathbb{C}}^{1 \times {N}}}$. In particular, we assume that the $K$ users are grouped into $M$ clusters, where $1 \le M \le K$ is an integer which can be optimized. For example, $M=1$ means that all $K$ users are grouped into only one cluster, while $M=K$ indicates that each cluster contains only one user. Let ${{\mathcal{G}}_m},\forall m \in {\mathcal{M}} \triangleq \left\{ {1,2, \ldots ,M} \right\}$, denote the set of users in the $m$th cluster, where ${{\mathcal{G}}_1} \cup {{\mathcal{G}}_2} \cup  \ldots  \cup {{\mathcal{G}}_M} = {\mathcal{K}}$. The main purpose of user grouping is to employ NOMA transmission. This is similar to the conventional cluster-based NOMA, in which, however, each cluster has to contain at least two users. Let the binary variable $\alpha _{k,i}^m \in \left\{ {0,1} \right\},\forall k \ne i \in {{\mathcal{G}}_m},m \in {\mathcal{M}}$, specify the SIC decoding order of users $k$ and $i$ in the $m$th cluster. If $\alpha _{k,i}^m = 0$, it means that, for the $m$th cluster, user $k$ will carry out SIC to first decode user $i$'s signal and remove it from the received signal. Otherwise, we have $\alpha _{k,i}^m = 1$. Accordingly, we have the condition $\alpha _{k,i}^m + \alpha _{i,k}^m = 1$ since it is in general impossible to mutually carry out SIC at both users. \\
\indent Without loss of generality, we focus on the $k$th user who is assumed to be grouped into the $m$th cluster. The corresponding achievable communication rate is given by
\begin{align}\label{rate}
{R_{{\rm{DL}},k}^{{\rm{NGMA}}}}\! \!=\!\! {\log _2}\!\left( \!\!{1 \!+ \!\!\frac{{{{\left| {{\mathbf{h}}_k^H{{\mathbf{w}}_k}} \right|}^2}}}{{\sum\limits_{i \ne k,i \in {{\mathcal{G}}_m}}\!\!\!\!\!{\alpha _{k,i}^m{{\left| {{\mathbf{h}}_k^H{{\mathbf{w}}_i}} \right|}^2}}\! \! +\!\!\!\!\! \sum\limits_{j \in {\mathcal{K}}/{{\mathcal{G}}_m}}\!\!\!\!\!{{{\left| {{\mathbf{h}}_k^H{{\mathbf{w}}_j}} \right|}^2}}\! \! +\! \sigma _k^2}}} \right).
\end{align}
\begin{table*}[!t]\small
\caption{Summary of NGMA and existing schemes for downlink transmission.}
\vspace{-0.2cm}
\begin{center}
\centering
\resizebox{\textwidth}{!}{
\begin{tabular}{!{\vrule width1.0pt}l!{\vrule width1.0pt}l!{\vrule width1.0pt}l!{\vrule width1.0pt}l!{\vrule width1.0pt}l!{\vrule width1.0pt}}
\Xhline{1.0pt}
\centering
\makecell[c]{}  & \makecell[c]{\textbf{NGMA}} &\makecell[c]{\textbf{SDMA}\\ ($M=K$)} & \makecell[c]{\textbf{BB-NOMA}\\($M=1$)} & \makecell[c]{\textbf{CB-NOMA}\\($1 < M < K, \left| {{{\mathcal{G}}_m}} \right| \ge 2, {\overline {\mathbf{w}} _k} \triangleq {\overline {\mathbf{w}} _{{{\mathcal{G}}_m}}},\forall k \in {{\mathcal{G}}_m}$)}\\
\Xhline{1.0pt}
\centering
\makecell[c]{Beamformer, ${{{\mathbf{w}}_k}}$} & \makecell[c]{$\surd$} & \makecell[c]{$\surd$} & \makecell[c]{$\surd$} & \makecell[c]{$\surd$} \\
\hline
\centering
\makecell[c]{User grouping, ${{{\mathcal{G}}_m}}$}& \makecell[c]{$\surd$}  &\makecell[c]{$\times$}  &\makecell[c]{$\times$} & \makecell[c]{$\surd$}\\
\hline
\centering
\makecell[c]{User ordering, ${\alpha _{k,i}^m}$}& \makecell[c]{$\surd$}   & \makecell[c]{$\times$}    & \makecell[c]{$\surd$} &\makecell[c]{$\surd$} \\
\hline
\centering
\makecell[c]{Relationship}&   &   \makecell[c]{${\mathcal{R}_{\rm{DL}}^{{\rm{NGMA}}}} \supseteq {\mathcal{R}_{\rm{DL}}^{{\rm{SDMA}}}}$}  & \makecell[c]{${\mathcal{R}_{\rm{DL}}^{{\rm{NGMA}}}} \supseteq {\mathcal{R}_{\rm{DL}}^{{\rm{BB-NOMA}}}}$} & \makecell[c]{${\mathcal{R}_{\rm{DL}}^{{\rm{NGMA}}}} \supseteq {\mathcal{R}_{\rm{DL}}^{{\rm{CB-NOMA}}}}$} \\
\Xhline{1.0pt}
\end{tabular}
}
\end{center}
\label{table:structure}
\end{table*}
\noindent The SIC condition for user $k$ to decode the signal of user $i$ in the $m$th cluster (i.e., $\alpha _{k,i}^m = 0$) is given by
\begin{align}\label{SIC NGMA DL}
{R_{{\rm{DL}},i \to k}^{{\rm{NGMA}}}} \ge {R_{{\rm{DL}},i \to i}^{{\rm{NGMA}}}},
\end{align}
where
\[{R_{{\rm{DL}},i \to k}^{{\rm{NGMA}}}} \!\!=\!\! {\log _2}\!\left( \!\!{1 \!+ \!\frac{{{{\left| {{\mathbf{h}}_k^H{{\mathbf{w}}_i}} \right|}^2}}}{{\sum\limits_{l \ne i,l \in {{\mathcal{G}}_m}}\!\!\!\!\!{\alpha _{l,i}^m{{\left| {{\mathbf{h}}_k^H{{\mathbf{w}}_l}} \right|}^2}}\!\!  +\!\!\!\!\! \sum\limits_{j \in {\mathcal{K}}/{{\mathcal{G}}_m}}\!\!\!\!\!{{{\left| {{\mathbf{h}}_k^H{{\mathbf{w}}_j}} \right|}^2}}\!  \!+\! \sigma _k^2}}} \right)\!,\]
and
\[{R_{{\rm{DL}},i \to i}^{{\rm{NGMA}}}} \!\!=\!\! {\log _2}\!\left( \!\!{1 \!+ \!\frac{{{{\left| {{\mathbf{h}}_i^H{{\mathbf{w}}_i}} \right|}^2}}}{{\sum\limits_{l \ne i,l \in {{\mathcal{G}}_m}}\!\!\!\!\!{\alpha _{l,i}^m{{\left| {{\mathbf{h}}_i^H{{\mathbf{w}}_l}} \right|}^2}}\! \! +\!\!\!\!\!\! \sum\limits_{j \in {\mathcal{K}}/{{\mathcal{G}}_m}}\!\!\!\!\!{{{\left| {{\mathbf{h}}_i^H{{\mathbf{w}}_j}} \right|}^2}}\! \! +\! \sigma _i^2}}} \right)\!.\]
Based on the above signal model, in the following, we will discuss some special cases of the proposed unified NGMA framework to further demonstrate its ability to provide flexible transmission schemes.\\
\indent \emph{\textbf{Special case 1 (SDMA)}}: In this case, we assume that $M = K$, i.e., each cluster contains only-one user and thus NOMA is not employed. Then, the achievable rate of user $k$ is given by
\begin{align}\label{rate SC1}
R_{{\rm{DL}},k}^{{\rm{SDMA}}} = {\log _2}\left( {1 + \frac{{{{\left| {{\mathbf{h}}_k^H{{\mathbf{w}}_k}} \right|}^2}}}{{\sum\limits_{j \ne k,j \in {\mathcal{K}}} {{{\left| {{\mathbf{h}}_k^H{{\mathbf{w}}_j}} \right|}^2}}  + \sigma _k^2}}} \right),
\end{align}
which corresponds to standard SDMA as shown in the top left of Fig. \ref{Downlink}. We note that SDMA is only applicable in the underloaded/critically loaded regime (i.e., $K \le N$) with sufficient spatial DoFs to design beamformers such that ${\left| {{\mathbf{h}}_k^H{{\mathbf{w}}_j}} \right|^2} \to 0,\forall k \ne j \in {\mathcal{K}}$.\\
\indent \emph{\textbf{Special case 2 (Beamformer-based NOMA)}}: In this case, we assume that $M = 1$, i.e., all users are grouped into only one cluster and NOMA is employed. Then, the achievable rate of user $k$ is given by
\begin{align}\label{rate SC2}
R_{{\rm{DL}},k}^{{\rm{BB-NOMA}}} = {\log _2}\left( {1 + \frac{{{{\left| {{\mathbf{h}}_k^H{{\mathbf{w}}_k}} \right|}^2}}}{{\sum\limits_{i \ne k,i \in {{\mathcal{K}}}}  {\alpha _{k,i}}{{\left| {{\mathbf{h}}_k^H{{\mathbf{w}}_i}} \right|}^2} + \sigma _k^2}}} \right),
\end{align}
which corresponds to standard beamformer-based NOMA as shown in the top middle of Fig. \ref{Downlink}. Beamformer-based NOMA is mainly beneficial in underloaded/critically loaded systems with strongly correlated channels and moderately overloaded systems (e.g., $N < K < 2N$). \\
\indent \emph{\textbf{Special case 3 (Cluster-based NOMA)}}: In this case, we assume that $1 < M < K$ and each cluster contains at least two users, i.e., $\left| {{{\mathcal{G}}_m}} \right| \ge 2,\forall m \in {\mathcal{M}}$. Moreover, for each cluster, we have ${\overline {\mathbf{w}} _k} \triangleq {\overline {\mathbf{w}} _{{{\mathcal{G}}_m}}},\forall k \in {{\mathcal{G}}_m}$. Accordingly, the achievable rate of user $k$ in the $m$th cluster is given by \eqref{rate SC3} shown at the top of the next page,
\begin{figure*}[!t]\normalsize
\begin{align}\label{rate SC3}
R_k^{{\rm{CB-NOMA}}} =  {\log _2}\left( {1 + \frac{{{p_k}{{\left| {{\mathbf{h}}_k^H{{\overline {\mathbf{w}} }_{{{\mathcal{G}}_m}}}} \right|}^2}}}{{{{\left| {{\mathbf{h}}_k^H{{\overline {\mathbf{w}} }_{{{\mathcal{G}}_m}}}} \right|}^2}\sum\limits_{i \ne k,i \in {{{\mathcal{G}}}_m}}  \alpha _{k,i}^m{p_i} + \sum\limits_{n \ne m,n \in {{\mathcal{M}}}} {{{\left| {{\mathbf{h}}_k^H{{\overline {\mathbf{w}} }_{{{\mathcal{K}}_n}}}} \right|}^2}} \sum\limits_{j \in {{{\mathcal{G}}}_n}}  {p_j} + \sigma _k^2}}} \right).
\end{align}
\hrulefill \vspace*{0pt}
\end{figure*}
which corresponds to standard cluster-based NOMA as shown in the top right of Fig. \ref{Downlink}. Cluster-based NOMA is practically suitable for extremely overloaded systems, i.e., $K \gg N$. \\
\indent Based on the above considerations, it can be observed that the proposed unified NGMA framework includes the existing transmission schemes as special cases and provides enhanced DoFs by combining the benefits of the both multiple antenna and NOMA techniques. This is because, regardless of whether the system is underloaded/critically loaded or overloaded, the proposed NGMA framework can find a flexible transmission strategy instead of only switching between existing schemes. Table \ref{table:structure} summarizes the optimization variables of the proposed unified NGMA framework and the existing transmission schemes and identifies their relationships for downlink transmission, where ``BB-NOMA'' and ``CB-NOMA'' refer to beamformer-based NOMA and cluster-based NOMA, respectively, and ${{{\mathcal{R}}}_{\rm{DL}}^{\rm{X}}}$ denotes the achievable rate region of each scheme in the downlink case.

\subsection{A Unified Framework for NGMA: Uplink Case}
\begin{figure*}[t!]
    \begin{center}
        \includegraphics[width=6in]{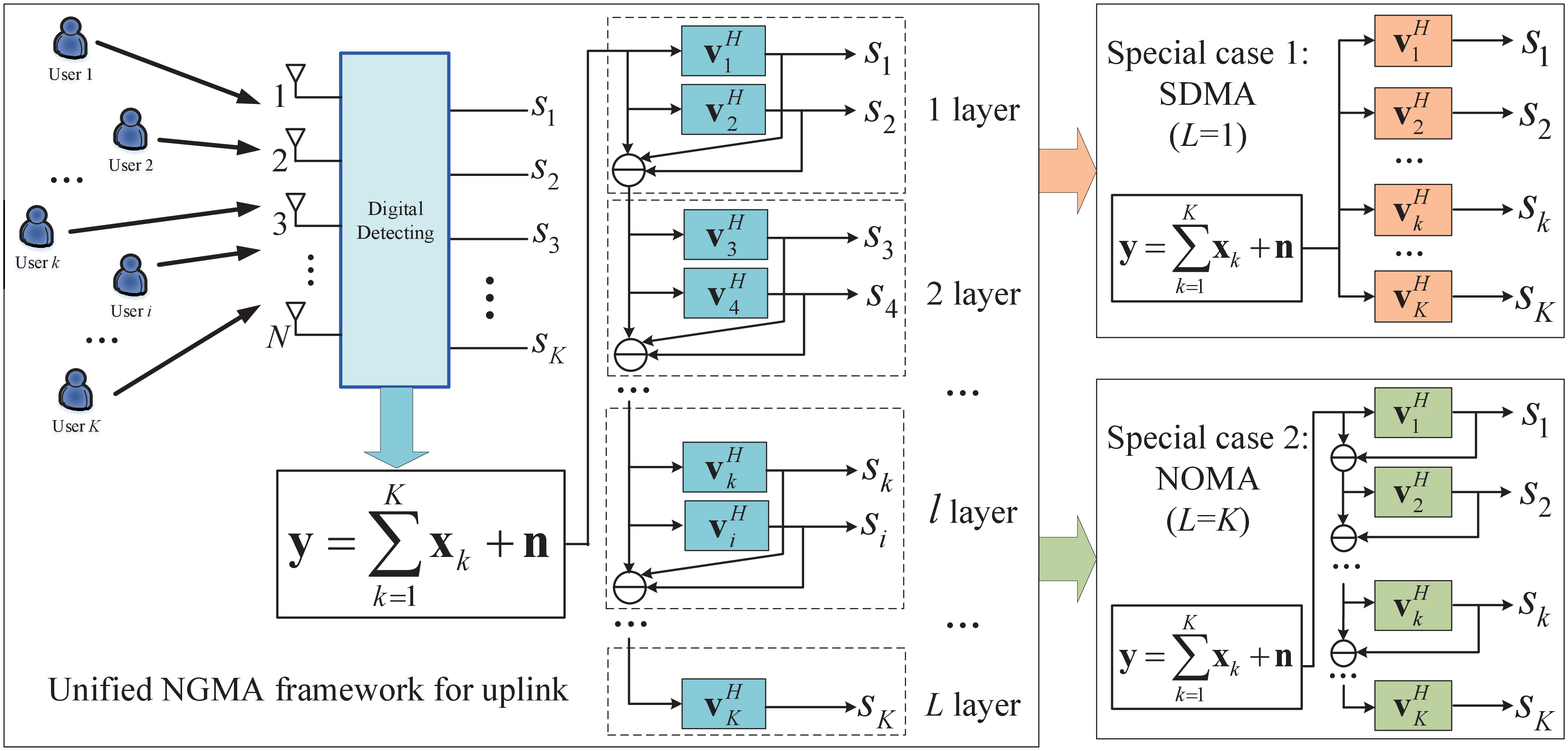}
        \caption{Illustration of the proposed multi-antenna and NOMA-based NGMA framework for uplink transmission.}
        \label{Uplink}
    \end{center}
    \end{figure*}

As shown in Fig. \ref{Uplink}, in this subsection, we consider the uplink of a typical SIMO multi-user communication system, where $K>1$ single-antenna users transmit their information to an $N$-antenna BS. The information symbol transmitted by user $k$ is denoted by $s_k$ and the corresponding transmit power is denoted by $p_k$. Let ${\mathbf{h}}_k \in {{\mathbb{C}}^{{N}\times 1}}$ denote the uplink channel coefficient between user $k$ and the BS. Therefore, the received signal at the BS is given by
\begin{align}\label{received signal uplink}
\begin{gathered}
  {\mathbf{y}} = \sum\nolimits_{k = 1}^K {{{\mathbf{h}}_k}\sqrt {{p_k}} {s_k}}  + {\mathbf{n}}, \hfill \\
  \;\;\; = \sum\nolimits_{k = 1}^K {{{\mathbf{x}}_k}}  + {\mathbf{n}}, \hfill \\
\end{gathered}
\end{align}
where ${{\mathbf{x}}_k} \triangleq {{\mathbf{h}}_k}\sqrt {{p_k}} {s_k}$ denotes the received signal at the BS from user $k$, and ${{\mathbf{n}}} \sim {\mathcal{C}\mathcal{N}}\left( {0,\sigma ^2{{\mathbf{I}}_{{N}}}} \right)$ denotes the AWGN at the BS.\\
\indent Different from the downlink case, where each user only aims to decode its intended signal based on its received signal, the BS in the uplink case has to decode the received $K$ information streams, i.e., ${\mathbf{s}} = {\left[ {{s_1},{s_2}, \ldots ,{s_K}} \right]^T}$. In other words, from the perspective of the BS, all signals are intended signals. Let ${{\mathbf{v}}_k^H}\in {{\mathbb{C}}^{1 \times {N}}}$ denote the normalized detection vector employed at the BS for decoding $s_k$ and the corresponding output signal is given by
\begin{align}\label{detected signal}
\begin{gathered}
  {y_{k}} = {\mathbf{v}}_k^H{{\mathbf{y}}} \hfill \\
   \;\;\;\;\;= \underbrace {{\mathbf{v}}_k^H{{\mathbf{h}}_k}\sqrt {{p_k}} {s_k}}_{{\rm{desired\;signal}}} + \underbrace {{\mathbf{v}}_k^H\left( {\sum\nolimits_{i \ne k} {{{\mathbf{h}}_i}\sqrt {{p_i}} {s_i}} } \right)}_{{\rm{inter - user\;interference}}} + {\mathbf{v}}_k^H{{\mathbf{n}}}. \hfill \\
\end{gathered}
\end{align}
Before introducing our proposed unified framework for NGMA for uplink transmission, we first review the existing SDMA-based and NOMA-based uplink signal detection schemes and discuss their benefits and drawbacks.
\subsubsection{\textbf{SDMA (Parallel detection)}} Similar to SDMA in the downlink case, the BS directly decodes $s_k$ in the presence of inter-user interference. The corresponding achievable communication rate of $s_k$ is given by
\begin{align}\label{SINR DD}
R_{{\rm{UL}},k}^{{\rm{SDMA}}} \!\!=\!\! {\log _2}\!\left( {1 + \frac{{{{\left| {{\mathbf{v}}_k^H{{\mathbf{h}}_k}} \right|}^2}{p_k}}}{{\underbrace {\sum\limits_{i \ne k} {{{\left| {{\mathbf{v}}_k^H{{\mathbf{h}}_i}} \right|}^2}{p_i}} }_{{\rm{interference\;from\;all\;the\;other\;users}}} \!\!\!+ {\sigma ^2}}}} \!\right).
\end{align}
To guarantee the detection performance, ${\mathbf{v}}_k^H$ has to be designed to completely mitigate the interference or to mitigate it to a large extend, e.g., $\sum\limits_{i \ne k} {{{\left| {{\mathbf{v}}_k^H{{\mathbf{h}}_i}} \right|}^2}}  = 0$ or $\sum\limits_{i \ne k} {{{\left| {{\mathbf{v}}_k^H{{\mathbf{h}}_i}} \right|}^2}}  \approx 0$, which is only possible when $K \le N$. Therefore, if the system is underloaded and the users' channels are sufficiently uncorrelated, SDMA is promising. Moreover, employing SDMA to detect all received information streams can be regarded as a parallel detection process, i.e., each information stream is separately detected using one unique detection vector. Such a parallel processing implies low latency, which is an important property for future wireless communication systems. However, when the system is overloaded or the users' channels are strongly correlated, SDMA cannot achieve high performance due to the ineffective interference mitigation, similar to the downlink case.
\subsubsection{\textbf{NOMA (Serial detection)}} Another typical uplink detection scheme is SIC, where the BS detects all the information streams one by one as is typical for NOMA. In particular, the BS first detects one selected signal and subtracts it from the received signal, then decoding other signals without the interference from this already decoded signal. The corresponding achievable communication rate for $s_k$ is given by
\begin{align}\label{SINR SNOMA}
R_{{\rm{UL}},k}^{{\rm{NOMA}}}\!\! =\!\! {\log _2}\!\left( {1 + \frac{{{{\left| {{\mathbf{v}}_k^H{{\mathbf{h}}_k}} \right|}^2}{p_k}}}{{\underbrace {\sum\limits_{{\beta _i} > {\beta _k}} {{{\left| {{\mathbf{v}}_k^H{{\mathbf{h}}_i}} \right|}^2}{p_i}} }_{{\rm{interference\;from\;the\;undetected\;users}}} \!\!\!\!+ {\sigma ^2}}}} \!\right),
\end{align}
where ${\beta _k} \in {\mathcal{K}} \triangleq \left\{ {1,2, \ldots ,K} \right\}$ represents the decoding order of $s_k$ and ${\beta _k} \ne {\beta _i},\forall k \ne i \in {\mathcal{K}}$. For uplink NOMA, the SIC decoding condition needed in the downlink case does not exist since there is only one receiver and all signals are desired. As a result, the BS can decode the received signals in any decoding order. However, the decoding orders generally determine the achievable communication rate of each information stream, which is indeed expected since the later the signal is decoded, the less interference it will suffer from. Therefore, the uplink NOMA SIC decoding order can be optimized to satisfy a given rate requirement for each user. Similar to downlink NOMA transmission, uplink NOMA is mainly beneficial for overloaded systems, where the spatial DoFs are insufficient for mitigating inter-user interference and SIC is necessary. However, uplink NOMA can also be applied in underloaded/critically loaded systems and its performance will be no worse than that of SDMA in terms of communication rate, which is different from downlink multi-antenna transmission, where NOMA may perform worse than SDMA in underloaded /critically loaded systems (see Section VII.A.1.). This is because using the same detection vectors, NOMA can employ SIC to further cancel inter-user interference that may be not completely mitigated for SDMA. However, the serial detection in NOMA will increase the latency and requires a higher computational complexity because of user ordering, especially for large $K$, which constitutes the main drawback of NOMA for uplink transmission.

\subsubsection{\textbf{A unified NGMA framework for uplink transmission}} Based on the above discussion, it can be concluded that for the uplink case, NOMA achieves a higher rate while SDMA is preferable in terms of latency and complexity. However, given the uncertainty of wireless environments as well as the heterogenous QoS and latency requirements, we should not rely on either NOMA or SDMA alone to tackle the resulting challenges. Therefore, in the following, we propose a unified NGMA framework for uplink transmission to combine the advantages of SDMA and NOMA.\\
\indent We still consider the aforementioned uplink system model. Instead of decoding signals in a pure parallel or serial manner, we divide the detection of the $K$ information streams into $L$ layers, where $1 \le L \le K$, as shown in Fig. \ref{Uplink}. In particular, within each layer, some properly selected information streams are decoded in a parallel manner, as in SDMA. Once these signals are decoded, they are subtracted from the currently undetected signals. As a result, the signals in the later layers can be detected without the interference from the signals which have been already decoded in the earlier layers. In other words, for the proposed unified framework for uplink transmission, SDMA is employed within each layer and NOMA is employed between the layers. Without loss of generality, let ${{\mathcal{U}}_l},\forall l \in {\mathcal{L}} \triangleq \left\{ {1,2, \ldots ,L} \right\}$, denote the set of information streams to be decoded in the $l$th layer, where ${{\mathcal{U}}_1} \cup {{\mathcal{U}}_2} \cup  \ldots  \cup {{\mathcal{U}}_L} = {\mathcal{K}}$. Assume that $s_k$ is decoded in the $l$th layer employing detection vector ${\mathbf{v}}_k^H$, the achievable communication rate of $s_k$ in the proposed NGMA framework can be expressed as \eqref{SINR NGMA} shown at the top of the this page.
\begin{figure*}[!t]\normalsize
\begin{align}\label{SINR NGMA}
R_{{\rm{UL}},k}^{{\rm{NGMA}}} = {\log _2}\left( {1 + \frac{{{{\left| {{\mathbf{v}}_k^H{{\mathbf{h}}_k}} \right|}^2}{p_k}}}{{\underbrace {\sum\limits_{i \ne k,i \in {{{\mathcal{U}}}_l}}  {{\left| {{\mathbf{v}}_k^H{{\mathbf{h}}_i}} \right|}^2}{p_i}}_{{\rm{inerference\;from\;signals\;in\;the\;current\;layer}}} + \underbrace {\sum\limits_{j \in {{{\mathcal{U}}}_n},n > l \in {\mathcal{L}}}  {{\left| {{\mathbf{v}}_k^H{{\mathbf{h}}_j}} \right|}^2}{p_j}}_{{\rm{inerference\;from\;signals\;in\;the\;later\;layers}}} + {\sigma ^2}}}} \right).
\end{align}
\hrulefill \vspace*{0pt}
\end{figure*}
Based on \eqref{SINR NGMA}, we consider the following two special cases.\\
\indent \emph{\textbf{Special case 1}}: For $L=1$, there is only one layer, in which all $K$ information streams have to be decoded, as shown in the top right of Fig. \ref{Uplink}. In this case, the $K$ information streams are decoded in a pure parallel manner by treating the other signals as interference, i.e., SDMA is used. Then, \eqref{SINR NGMA} can be simplified to \eqref{SINR DD}.\\
\indent \emph{\textbf{Special case 2}}: For $L=K$, there is only one signal to be decoded in each layer, as shown in the bottom right of Fig. \ref{Uplink}. In this case, the $K$ information streams are decoded in a pure serial manner with the interference from the previously decoded signals subtracted, i.e., NOMA is used. Then, \eqref{SINR NGMA} can be simplified to \eqref{SINR SNOMA}.\\
\indent Therefore, the proposed unified uplink NGMA framework integrates both SDMA and NOMA and provides more DoFs for signal detection. Table \ref{table:structure2} summarizes the optimization variables of the proposed NGMA framework and the existing transmission schemes as well as identifies their characteristics and relationships, where ${{{\mathcal{R}}}_{\rm{UL}}^{\rm{X}}}$ denotes the achievable rate region of each scheme in the uplink case.
\begin{table*}[!t]\small
\caption{Summary of NGMA and existing schemes for uplink transmission.}
\vspace{-0.2cm}
\begin{center}
\centering
\begin{tabular}{!{\vrule width1.0pt}l!{\vrule width1.0pt}l!{\vrule width1.0pt}l!{\vrule width1.0pt}l!{\vrule width1.0pt}}
\Xhline{1.0pt}
\centering
\makecell[c]{}  & \makecell[c]{\textbf{NGMA}} &\makecell[c]{\textbf{SDMA}\\ ($L=1$)} & \makecell[c]{\textbf{NOMA}\\($L=K$)} \\
\Xhline{1.0pt}
\centering
\makecell[c]{Detection vector} & \makecell[c]{$\surd$} & \makecell[c]{$\surd$} & \makecell[c]{$\surd$}  \\
\hline
\centering
\makecell[c]{Decoding order}& \makecell[c]{$\surd$}  &\makecell[c]{$\times$}  &\makecell[c]{$\surd$} \\
\hline
\centering
\makecell[c]{Layer classification}& \makecell[c]{$\surd$}   & \makecell[c]{$\times$}    & \makecell[c]{$\times$} \\
\hline
\centering
\makecell[c]{Characteristics}& \makecell[c]{High flexibility\\Good rate-versus-latency tradeoff}   & \makecell[c]{Performance loss\\Low complexity and latency}    & \makecell[c]{High rate gain\\High latency} \\
\hline
\centering
\makecell[c]{Relationship}&   \multicolumn{3}{c|}{\makecell[c]{${\mathcal{R}}_{{\rm{NGMA}}}^{{\rm{UL}}} = {\mathcal{R}}_{{\rm{NOMA}}}^{{\rm{UL}}} \supseteq {\mathcal{R}}_{{\rm{SDMA}}}^{{\rm{UL}}}$}} \\
\Xhline{1.0pt}
\end{tabular}
\end{center}
\label{table:structure2}
\end{table*}
\subsection{Discussion and Outlook}
Based on the multi-antenna and NOMA techniques, a unified NGMA framework is proposed for both downlink and uplink transmission, which is shown to be able to generalize the existing transmission schemes and to go beyond them. This opens up new research opportunities and directions, two of which are exemplified below.

\subsubsection{Joint user grouping and beamformer design} It can be observed that user grouping plays an important role in the proposed unified NGMA framework, which further determines the beamformer design and/or resource allocation. However, the user grouping design introduces binary variables, which makes the resulting optimization problem non-trivial to solve. To address this issue, the previously introduced matching theory may be an effective tool, which deserves further investigation. Furthermore, when encountering dynamic environments, i.e., the users frequently join and leave the wireless network, the resulting optimization problem under the proposed unified NGMA framework will be quite challenging. In this case, ML tools can be employed to learn from the environment and provide a high-quality solution for guiding user grouping, beamformer design, and resource allocation. This constitutes another interesting topic for future work.

\subsubsection{Extensions to multiple-antenna user case} The proposed unified NGMA framework assumes that the users are equipped with a single-antenna and each user only has one intended data stream. However, for next generation wireless networks, both the BSs and users will have multiple antennas and each user will have to support multiple data streams. How to extend the proposed NGMA framework to general multi-user MIMO commination systems is an important research direction. It is also worth noting that communication systems with multi-antenna users are more likely to be overloaded as compared to systems with single-antenna users. Therefore, the resulting joint design of the transmit and receive beamformers become much more challenging and requires higher computational complexity. To this end, some research efforts have been devoted to exploiting the benefits of both NOMA and multiple antenna techniques in MIMO transmission. In particular, the authors of \cite{8558721} proposed a transmission framework for a downlink two-user MIMO communication system, which exploited the generalized singular value decomposition (GSVD) and NOMA. Multiple data streams were transmitted to both users in a hybrid NOMA/OMA manner based on properly designed precoders and decoders, thus enabling a more efficient use of the available spatial DoFs as compared to conventional MIMO-OMA and MIMO-NOMA. As a further step, the authors of \cite{9250605} proposed a User-Assisted Simultaneous Diagonalization (UA-SD) based MIMO-NOMA transmission scheme. By relying on low-complexity self-interference cancellation at the users, the proposed simultaneous diagonalization scheme in \cite{9250605} overcomes not only the drawbacks of conventional ZF and Block Diagonalization (BD) methods, which are only applicable in the underloaded/critically loaded regime, but also the drawbacks of the GSVD method \cite{8558721}, which may be ineffective when the number of BS antennas and the number of user antennas are comparable. Moreover, the authors of \cite{9321706} proposed a Simultaneous Triangularization (ST) based MIMO-NOMA scheme for both uplink and downlink transmission, which avoids channel inversion at transmitter and receiver as is required for SD based MIMO-NOMA~\cite{8558721,9250605}, thus improving the ergodic rate performance. Although the authors of \cite{8558721,9250605,9321706} mainly focused on the two-user case, the proposed schemes provide useful guidance for developing NGMA schemes for general multi-user MIMO commination systems, which is an important topic for future research.
\section{From NOMA to NGMA: Implementation Challenges and Future Work}
As the research on NGMA is still in a very early stage, besides the research directions that have already been discussed in the above sections, there are several implementation challenges to be tackled. In this section, we identify some of them to motivate future work on NGMA.
\subsection{Modulation and Detection Design}
Most existing works on NOMA assume that the superimposed signals of the different users are taken from an ideal Gaussian codebook, and thus employed the achievable rate as performance metric. However, in practical implementations, the superimposed signal may comprise signals using different modulation schemes (e.g., Binary Phase Shift Keying (BPSK) and Quadrature Amplitude Modulation (QAM))~\cite{6204010}. To ensure that the theoretically achievable rates can be realized in practice, efficient modulation and detection schemes have to be developed, which is a non-trivial task. This is because different from OMA, where the modulation and detection of the symbols for different users is carried out independently via orthogonal resource blocks, in NOMA, the users have to demodulate their intended and/or other users' symbols based on the received superimposed signal. For example, erroneous demodulation of the weak user's symbol will inevitably lead to a higher error probability when demodulating a strong user's own symbol. Therefore, the modulation scheme and signal power allocation for the different NOMA users have to be judiciously designed to improve the performance of demodulation~\cite{16_CL_Choi,19_TCOM_Zhao,19_TVT_Yeon,20_CL_Assaf}. For instance, the author of \cite{16_CL_Choi} studied the power allocation problem for a downlink two-user NOMA system employing practical modulation schemes. The results revealed that the optimal power allocation depends on the modulation constellations employed by the two users. Allowing arbitrary QAM modulation orders at both users, the authors of \cite{20_CL_Assaf} characterized the exact Bit Error Rate (BER) of a downlink two-user NOMA system, where the power allocation to the two users was shown to have a high impact on the BER performance.\\
\indent In the above works, it was assumed that the modulation type and order of the NOMA users are known at the receivers for demodulation. However, in practice, this may not be the case, since the acquisition would require additional higher-layer signaling overhead. To address this issue, motivated by the excellent classification capabilities of ML tools, ML-empowered blind modulation detection schemes \cite{18_CL_Zhang} are promising solutions for demodulating superimposed signals in NGMA, and constitute an interesting research topic. On the other hand, given the multiple-antenna feature of next generation wireless networks, new emerging modulation schemes such as Spatial Modulation (SM)~\cite{11_CM_Renzo} and Index Modulation (IM)~\cite{16_CM_Basar} are also promising techniques for adoption in NGMA, and have drawn some initial research interest~\cite{19_JSTSP_Li,19_JSTSP_Yang,20_SPL_Tusha,20_WCL_Shahab} but require future research.

\subsection{Error Propagation Mitigation}
As explained in the previous subsection, SIC is the key technique to allow users to successively demodulate other users' symbols as well as their own symbols from the received superimposed signal. However, one of the major bottlenecks for implementing SIC in practice is error propagation. For example, once the strong user demodulates the weak user's symbol erroneously in the first stage, it will subtract the wrong remodulated symbol from the received superimposed signal, which results in residual interference \cite{MD} and degrades the performance in the following stage. Since next generation wireless networks are expected to be overloaded, the numbers of users in the same NOMA cluster may be much larger than two. Therefore, SIC error propagation may be severe in NGMA. This is because if the symbols of the users with lower decoding orders are not correctly demodulated, the error probability of the remaining users' symbols will increase.\\
\indent Most existing work on NOMA is based on the assumption of perfect SIC, i.e., the receivers are assumed to perfectly demodulate the signals and remove the corresponding interference in each SIC stage as long as the corresponding constraints on the rate and/or power allocation are satisfied. Nevertheless, perfect SIC generally cannot be achieved in practice. As a result, on the one hand, some researchers have begun to investigate the effect of imperfect SIC in NOMA based two-way relay systems~\cite{8316931}, single-cell multi-antenna NOMA systems \cite{8554298}, and multi-cell uplink NOMA systems~\cite{9146345}. Note that in these works, the residual interference caused by imperfect SIC was simply modeled as a linear function of the received signal power, which may not be accurate enough when practical modulation schemes are considered. Therefore, sophisticated mathematical models for characterizing and understanding the effect of imperfect SIC are an interesting but challenging research direction, which may have to be based on data collected in field test. On the other hand, several works have also developed possible methods for solving or alleviating error propagation issues, such as using multistage channel estimation~\cite{942508} and employing iterative SIC-aided receivers~\cite{5522468}. These proposed methods improve the accuracy of SIC at the cost of increasing complexity. The development of methods that improve SIC performance while striking a good performance-versus-complexity tradeoff is an important research topic for NGMA.
\subsection{Advanced Channel Estimation Techniques}
Compared to OMA, the acquisition of accurate CSI is more important for NOMA and NGMA~\cite{6204010}. On the one hand, as can be observed from Section VII, the user grouping and ordering schemes depend on the CSI of each user, namely the correlation of the channels of different users and the channel strength disparity. On the other hand, the performance of SIC is also affected by the accuracy of the CSI, since the CSI is needed for reconstruction of the decoded signal for subtraction. As a result, channel estimation errors will result not only in user grouping and ordering errors, but also decoding errors.\\
\indent Given the expected large numbers of connected users and equipped antennas in next generation wireless networks, channel estimation in NGMA will be very challenging. This is because the high-performance near-optimal channel estimation algorithms proposed for conventional systems will cause an unacceptable signalling overhead and a high computational complexity for NGMA, and thus cannot be applied. Therefore, more effective channel estimation methods tailored specifically for NGMA have to be developed to strike a good balance between complexity and performance, where both the conventional methods (e.g., compressive sensing) and ML tools can be exploited. This constitutes an interesting direction for future research. We expect that the channel estimation to be particularly challenging for higher frequency bands (e.g., mmWave or THz bands), frequency-selective and/or time-selective fading, and RIS-assisted wireless communication systems, which calls for new and efficient estimation methods. Moreover, as perfect CSI is difficult to obtain, NGMA transmission design based on imperfect CSI~\cite{7934461}, partial CSI~\cite{7361990}, and limited feedback CSI~\cite{7434594} (especially for frequency-selective and/or time-selective fading) is another important research topic.

\section{Conclusion}	
In this article, first, an overview of the development of NGMA was provided with a particular focus on the evolution of NOMA for satisfying the stringent challenges in next generation wireless networks. In particular, the fundamental information-theoretic capacity limits of NOMA were reviewed and new requirements and possible candidates for NGMA were discussed. Then, the state-of-the-art research contributions on NOMA were surveyed, including multi-antenna techniques for NOMA, promising future application scenarios for NOMA, the interplay between NOMA and other emerging physical layer techniques, and advanced mathematical optimization and machine learning tools for NOMA. Based on the existing foundation in the field of NOMA, a multi-antenna and NOMA-based unified framework for NGMA for downlink and uplink transmission was proposed. The proposed framework overcomes the shortcomings of conventional NOMA and SDMA and opens up new research opportunities. Finally, several practical implementation challenges for NGMA, which deserve further exploration in future research, have been discussed.

\bibliographystyle{IEEEtran}
\bibliography{mybib}

\end{document}